\documentclass[journal,10pt,twoside,twocolumn]{IEEEtran}
\usepackage{cite}
\usepackage[T1]{fontenc}
\usepackage{graphicx}
\usepackage{mathtools}
\usepackage{amssymb}
\usepackage{amsmath}
\usepackage{paralist}
\usepackage{subfigure}
\usepackage{booktabs} 
\usepackage{algorithm}
\usepackage{algorithmic}
\usepackage{amsthm}
\usepackage{multirow}
\usepackage{microtype}
\usepackage{hyperref}
\usepackage{balance}

\usepackage{amssymb}
\usepackage{amsfonts}
\usepackage{mathrsfs}
\usepackage{xspace}
\usepackage{bm}
\usepackage{upgreek}

\newcommand{\safemath}[2]{\newcommand{#1}{\ensuremath{#2}\xspace}}

\safemath{\bma}{\mathbf{a}}
\safemath{\bmb}{\mathbf{b}}
\safemath{\bmc}{\mathbf{c}}
\safemath{\bmd}{\mathbf{d}}
\safemath{\bme}{\mathbf{e}}
\safemath{\bmf}{\mathbf{f}}
\safemath{\bmg}{\mathbf{g}}
\safemath{\bmh}{\mathbf{h}}
\safemath{\bmi}{\mathbf{i}}
\safemath{\bmj}{\mathbf{j}}
\safemath{\bmk}{\mathbf{k}}
\safemath{\bml}{\mathbf{l}}
\safemath{\bmm}{\mathbf{m}}
\safemath{\bmn}{\mathbf{n}}
\safemath{\bmo}{\mathbf{o}}
\safemath{\bmp}{\mathbf{p}}
\safemath{\bmq}{\mathbf{q}}
\safemath{\bmr}{\mathbf{r}}
\safemath{\bms}{\mathbf{s}}
\safemath{\bmt}{\mathbf{t}}
\safemath{\bmu}{\mathbf{u}}
\safemath{\bmv}{\mathbf{v}}
\safemath{\bmw}{\mathbf{w}}
\safemath{\bmx}{\mathbf{x}}
\safemath{\bmy}{\mathbf{y}}
\safemath{\bmz}{\mathbf{z}}
\safemath{\bmzero}{\mathbf{0}}
\safemath{\bmone}{\mathbf{1}}

\bmdefine{\biad}{a}
\bmdefine{\bibd}{b}
\bmdefine{\bicd}{c}
\bmdefine{\bidd}{d}
\bmdefine{\bied}{e}
\bmdefine{\bifd}{f}
\bmdefine{\bigd}{g}
\bmdefine{\bihd}{h}
\bmdefine{\biid}{i}
\bmdefine{\bijd}{j}
\bmdefine{\bikd}{k}
\bmdefine{\bild}{l}
\bmdefine{\bimd}{m}
\bmdefine{\bind}{n}
\bmdefine{\biod}{o}
\bmdefine{\bipd}{p}
\bmdefine{\biqd}{q}
\bmdefine{\bird}{r}
\bmdefine{\bisd}{s}
\bmdefine{\bitd}{t}
\bmdefine{\biud}{u}
\bmdefine{\bivd}{v}
\bmdefine{\biwd}{w}
\bmdefine{\bixd}{x}
\bmdefine{\biyd}{y}
\bmdefine{\bizd}{z}

\bmdefine{\bixid}{\xi}
\bmdefine{\bilambdad}{\lambda}
\bmdefine{\bimud}{\mu}
\bmdefine{\bithetad}{\theta}
\bmdefine{\biphid}{\phi}
\bmdefine{\bideltad}{\delta}

\safemath{\bmia}{\biad}
\safemath{\bmib}{\bibd}
\safemath{\bmic}{\bicd}
\safemath{\bmid}{\bidd}
\safemath{\bmie}{\bied}
\safemath{\bmif}{\bifd}
\safemath{\bmig}{\bigd}
\safemath{\bmih}{\bihd}
\safemath{\bmii}{\biid}
\safemath{\bmij}{\bijd}
\safemath{\bmik}{\bikd}
\safemath{\bmil}{\bild}
\safemath{\bmim}{\bimd}
\safemath{\bmin}{\bind}
\safemath{\bmio}{\biod}
\safemath{\bmip}{\bipd}
\safemath{\bmiq}{\biqd}
\safemath{\bmir}{\bird}
\safemath{\bmis}{\bisd}
\safemath{\bmit}{\bitd}
\safemath{\bmiu}{\biud}
\safemath{\bmiv}{\bivd}
\safemath{\bmiw}{\biwd}
\safemath{\bmix}{\bixd}
\safemath{\bmiy}{\biyd}
\safemath{\bmiz}{\bizd}

\safemath{\bmxi}{\bixid}
\safemath{\bmlambda}{\bilambdad}
\safemath{\bmmu}{\bimud}
\safemath{\bmtheta}{\bithetad}
\safemath{\bmphi}{\biphid}
\safemath{\bmdelta}{\bideltad}

\safemath{\bA}{\mathbf{A}}
\safemath{\bB}{\mathbf{B}}
\safemath{\bC}{\mathbf{C}}
\safemath{\bD}{\mathbf{D}}
\safemath{\bE}{\mathbf{E}}
\safemath{\bF}{\mathbf{F}}
\safemath{\bG}{\mathbf{G}}
\safemath{\bH}{\mathbf{H}}
\safemath{\bI}{\mathbf{I}}
\safemath{\bJ}{\mathbf{J}}
\safemath{\bK}{\mathbf{K}}
\safemath{\bL}{\mathbf{L}}
\safemath{\bM}{\mathbf{M}}
\safemath{\bN}{\mathbf{N}}
\safemath{\bO}{\mathbf{O}}
\safemath{\bP}{\mathbf{P}}
\safemath{\bQ}{\mathbf{Q}}
\safemath{\bR}{\mathbf{R}}
\safemath{\bS}{\mathbf{S}}
\safemath{\bT}{\mathbf{T}}
\safemath{\bU}{\mathbf{U}}
\safemath{\bV}{\mathbf{V}}
\safemath{\bW}{\mathbf{W}}
\safemath{\bX}{\mathbf{X}}
\safemath{\bY}{\mathbf{Y}}
\safemath{\bZ}{\mathbf{Z}}

\safemath{\bZero}{\mathbf{0}}
\safemath{\bOne}{\mathbf{1}}
\safemath{\bDelta}{\mathbf{\Delta}}
\safemath{\bLambda}{\mathbf{\UpLambda}}
\safemath{\bPhi}{\mathbf{\Upphi}}
\safemath{\bSigma}{\mathbf{\Upsigma}}
\safemath{\bOmega}{\mathbf{\Upomega}}
\safemath{\bTheta}{\mathbf{\Uptheta}}

\bmdefine{\biAd}{A}
\bmdefine{\biBd}{B}
\bmdefine{\biCd}{C}
\bmdefine{\biDd}{D}
\bmdefine{\biEd}{E}
\bmdefine{\biFd}{F}
\bmdefine{\biGd}{G}
\bmdefine{\biHd}{H}
\bmdefine{\biId}{I}
\bmdefine{\biJd}{J}
\bmdefine{\biKd}{K}
\bmdefine{\biLd}{L}
\bmdefine{\biMd}{M}
\bmdefine{\biOd}{N}
\bmdefine{\biPd}{O}
\bmdefine{\biQd}{P}
\bmdefine{\biRd}{R}
\bmdefine{\biSd}{S}
\bmdefine{\biTd}{T}
\bmdefine{\biUd}{U}
\bmdefine{\biVd}{V}
\bmdefine{\biWd}{W}
\bmdefine{\biXd}{X}
\bmdefine{\biYd}{Y}
\bmdefine{\biZd}{Z}

\bmdefine{\biDelta}{\Delta}
\bmdefine{\biLambda}{\Lambda}
\bmdefine{\biPhi}{\Phi}
\bmdefine{\biSigma}{\Sigma}
\bmdefine{\biOmega}{\Omega}
\bmdefine{\biTheta}{\Theta}

\safemath{\bimA}{\biAd}
\safemath{\bimB}{\biBd}
\safemath{\bimC}{\biCd}
\safemath{\bimD}{\biDd}
\safemath{\bimE}{\biEd}
\safemath{\bimF}{\biFd}
\safemath{\bimG}{\biGd}
\safemath{\bimH}{\biHd}
\safemath{\bimI}{\biId}
\safemath{\bimJ}{\biJd}
\safemath{\bimK}{\biKd}
\safemath{\bimL}{\biLd}
\safemath{\bimM}{\biMd}
\safemath{\bimN}{\biNd}
\safemath{\bimO}{\biOd}
\safemath{\bimP}{\biPd}
\safemath{\bimQ}{\biQd}
\safemath{\bimR}{\biRd}
\safemath{\bimS}{\biSd}
\safemath{\bimT}{\biTd}
\safemath{\bimU}{\biUd}
\safemath{\bimV}{\biVd}
\safemath{\bimW}{\biWd}
\safemath{\bimX}{\biXd}
\safemath{\bimY}{\biYd}
\safemath{\bimZ}{\biZd}

\safemath{\bimDelta}{\biDelta}
\safemath{\bimLambda}{\biLambda}
\safemath{\bimPhi}{\biPhi}
\safemath{\bimSigma}{\biSigma}
\safemath{\bimOmega}{\biOmega}
\safemath{\bimTheta}{\biTheta}

\safemath{\setA}{\mathcal{A}}
\safemath{\setB}{\mathcal{B}}
\safemath{\setC}{\mathcal{C}}
\safemath{\setD}{\mathcal{D}}
\safemath{\setE}{\mathcal{E}}
\safemath{\setF}{\mathcal{F}}
\safemath{\setG}{\mathcal{G}}
\safemath{\setH}{\mathcal{H}}
\safemath{\setI}{\mathcal{I}}
\safemath{\setJ}{\mathcal{J}}
\safemath{\setK}{\mathcal{K}}
\safemath{\setL}{\mathcal{L}}
\safemath{\setM}{\mathcal{M}}
\safemath{\setN}{\mathcal{N}}
\safemath{\setO}{\mathcal{O}}
\safemath{\setP}{\mathcal{P}}
\safemath{\setQ}{\mathcal{Q}}
\safemath{\setR}{\mathcal{R}}
\safemath{\setS}{\mathcal{S}}
\safemath{\setT}{\mathcal{T}}
\safemath{\setU}{\mathcal{U}}
\safemath{\setV}{\mathcal{V}}
\safemath{\setW}{\mathcal{W}}
\safemath{\setX}{\mathcal{X}}
\safemath{\setY}{\mathcal{Y}}
\safemath{\setZ}{\mathcal{Z}}
\safemath{\emptySet}{\varnothing}

\safemath{\colA}{\mathscr{A}}
\safemath{\colB}{\mathscr{B}}
\safemath{\colC}{\mathscr{C}}
\safemath{\colD}{\mathscr{D}}
\safemath{\colE}{\mathscr{E}}
\safemath{\colF}{\mathscr{F}}
\safemath{\colG}{\mathscr{G}}
\safemath{\colH}{\mathscr{H}}
\safemath{\colI}{\mathscr{I}}
\safemath{\colJ}{\mathscr{J}}
\safemath{\colK}{\mathscr{K}}
\safemath{\colL}{\mathscr{L}}
\safemath{\colM}{\mathscr{M}}
\safemath{\colN}{\mathscr{N}}
\safemath{\colO}{\mathscr{O}}
\safemath{\colP}{\mathscr{P}}
\safemath{\colQ}{\mathscr{Q}}
\safemath{\colR}{\mathscr{R}}
\safemath{\colS}{\mathscr{S}}
\safemath{\colT}{\mathscr{T}}
\safemath{\colU}{\mathscr{U}}
\safemath{\colV}{\mathscr{V}}
\safemath{\colW}{\mathscr{W}}
\safemath{\colX}{\mathscr{X}}
\safemath{\colY}{\mathscr{Y}}
\safemath{\colZ}{\mathscr{Z}}

\safemath{\opA}{\mathbb{A}}
\safemath{\opB}{\mathbb{B}}
\safemath{\opC}{\mathbb{C}}
\safemath{\opD}{\mathbb{D}}
\safemath{\opE}{\mathbb{E}}
\safemath{\opF}{\mathbb{F}}
\safemath{\opG}{\mathbb{G}}
\safemath{\opH}{\mathbb{H}}
\safemath{\opI}{\mathbb{I}}
\safemath{\opJ}{\mathbb{J}}
\safemath{\opK}{\mathbb{K}}
\safemath{\opL}{\mathbb{L}}
\safemath{\opM}{\mathbb{M}}
\safemath{\opN}{\mathbb{N}}
\safemath{\opO}{\mathbb{O}}
\safemath{\opP}{\mathbb{P}}
\safemath{\opQ}{\mathbb{Q}}
\safemath{\opR}{\mathbb{R}}
\safemath{\opS}{\mathbb{S}}
\safemath{\opT}{\mathbb{T}}
\safemath{\opU}{\mathbb{U}}
\safemath{\opV}{\mathbb{V}}
\safemath{\opW}{\mathbb{W}}
\safemath{\opX}{\mathbb{X}}
\safemath{\opY}{\mathbb{Y}}
\safemath{\opZ}{\mathbb{Z}}
\safemath{\opZero}{\mathbb{O}}
\safemath{\identityop}{\opI}

\safemath{\veca}{\bma}
\safemath{\vecb}{\bmb}
\safemath{\vecc}{\bmc}
\safemath{\vecd}{\bmd}
\safemath{\vece}{\bme}
\safemath{\vecf}{\bmf}
\safemath{\vecg}{\bmg}
\safemath{\vech}{\bmh}
\safemath{\veci}{\bmi}
\safemath{\vecj}{\bmj}
\safemath{\veck}{\bmk}
\safemath{\vecl}{\bml}
\safemath{\vecm}{\bmm}
\safemath{\vecn}{\bmn}
\safemath{\veco}{\bmo}
\safemath{\vecp}{\bmp}
\safemath{\vecq}{\bmq}
\safemath{\vecr}{\bmr}
\safemath{\vecs}{\bms}
\safemath{\vect}{\bmt}
\safemath{\vecu}{\bmu}
\safemath{\vecv}{\bmv}
\safemath{\vecw}{\bmw}
\safemath{\vecx}{\bmx}
\safemath{\vecy}{\bmy}
\safemath{\vecz}{\bmz}

\safemath{\veczero}{\bmzero}
\safemath{\vecone}{\bmone}
\safemath{\vecxi}{\bmxi}
\safemath{\veclambda}{\bmlambda}
\safemath{\vecmu}{\bmmu}
\safemath{\vectheta}{\bmtheta}
\safemath{\vecphi}{\bmphi}
\safemath{\vecdelta}{\bmdelta}

\safemath{\matA}{\bA}
\safemath{\matB}{\bB}
\safemath{\matC}{\bC}
\safemath{\matD}{\bD}
\safemath{\matE}{\bE}
\safemath{\matF}{\bF}
\safemath{\matG}{\bG}
\safemath{\matH}{\bH}
\safemath{\matI}{\bI}
\safemath{\matJ}{\bJ}
\safemath{\matK}{\bK}
\safemath{\matL}{\bL}
\safemath{\matM}{\bM}
\safemath{\matN}{\bN}
\safemath{\matO}{\bO}
\safemath{\matP}{\bP}
\safemath{\matQ}{\bQ}
\safemath{\matR}{\bR}
\safemath{\matS}{\bS}
\safemath{\matT}{\bT}
\safemath{\matU}{\bU}
\safemath{\matV}{\bV}
\safemath{\matW}{\bW}
\safemath{\matX}{\bX}
\safemath{\matY}{\bY}
\safemath{\matZ}{\bZ}
\safemath{\matzero}{\bmzero}

\safemath{\matDelta}{\bDelta}
\safemath{\matLambda}{\bLambda}
\safemath{\matPhi}{\bPhi}
\safemath{\matSigma}{\bSigma}
\safemath{\matOmega}{\bOmega}
\safemath{\matTheta}{\bTheta}

\safemath{\matidentity}{\matI}
\safemath{\matone}{\matO}

\safemath{\rnda}{A}
\safemath{\rndb}{B}
\safemath{\rndc}{C}
\safemath{\rndd}{D}
\safemath{\rnde}{E}
\safemath{\rndf}{F}
\safemath{\rndg}{G}
\safemath{\rndh}{H}
\safemath{\rndi}{I}
\safemath{\rndj}{J}
\safemath{\rndk}{K}
\safemath{\rndl}{L}
\safemath{\rndm}{M}
\safemath{\rndn}{N}
\safemath{\rndo}{O}
\safemath{\rndp}{P}
\safemath{\rndq}{Q}
\safemath{\rndr}{R}
\safemath{\rnds}{S}
\safemath{\rndt}{T}
\safemath{\rndu}{U}
\safemath{\rndv}{V}
\safemath{\rndw}{W}
\safemath{\rndx}{X}
\safemath{\rndy}{Y}
\safemath{\rndz}{Z}

\safemath{\rveca}{\bimA}
\safemath{\rvecb}{\bimB}
\safemath{\rvecc}{\bimC}
\safemath{\rvecd}{\bimD}
\safemath{\rvece}{\bimE}
\safemath{\rvecf}{\bimF}
\safemath{\rvecg}{\bimG}
\safemath{\rvech}{\bimH}
\safemath{\rveci}{\bimI}
\safemath{\rvecj}{\bimJ}
\safemath{\rveck}{\bimK}
\safemath{\rvecl}{\bimL}
\safemath{\rvecm}{\bimM}
\safemath{\rvecn}{\bimN}
\safemath{\rveco}{\bomO}
\safemath{\rvecp}{\bimP}
\safemath{\rvecq}{\bimQ}
\safemath{\rvecr}{\bimR}
\safemath{\rvecs}{\bimS}
\safemath{\rvect}{\bimT}
\safemath{\rvecu}{\bimU}
\safemath{\rvecv}{\bimV}
\safemath{\rvecw}{\bimW}
\safemath{\rvecx}{\bimX}
\safemath{\rvecy}{\bimY}
\safemath{\rvecz}{\bimZ}

\safemath{\rvecxi}{\bmxi}
\safemath{\rveclambda}{\bmlambda}
\safemath{\rvecmu}{\bmmu}
\safemath{\rvectheta}{\bmtheta}
\safemath{\rvecphi}{\bmphi}

\safemath{\rmatA}{\bimA}
\safemath{\rmatB}{\bimB}
\safemath{\rmatC}{\bimC}
\safemath{\rmatD}{\bimD}
\safemath{\rmatE}{\bimE}
\safemath{\rmatF}{\bimF}
\safemath{\rmatG}{\bimG}
\safemath{\rmatH}{\bimH}
\safemath{\rmatI}{\bimI}
\safemath{\rmatJ}{\bimJ}
\safemath{\rmatK}{\bimK}
\safemath{\rmatL}{\bimL}
\safemath{\rmatM}{\bimM}
\safemath{\rmatN}{\bimN}
\safemath{\rmatO}{\bimO}
\safemath{\rmatP}{\bimP}
\safemath{\rmatQ}{\bimQ}
\safemath{\rmatR}{\bimR}
\safemath{\rmatS}{\bimS}
\safemath{\rmatT}{\bimT}
\safemath{\rmatU}{\bimU}
\safemath{\rmatV}{\bimV}
\safemath{\rmatW}{\bimW}
\safemath{\rmatX}{\bimX}
\safemath{\rmatY}{\bimY}
\safemath{\rmatZ}{\bimZ}

\safemath{\rmatDelta}{\bimDelta}
\safemath{\rmatLambda}{\bimLambda}
\safemath{\rmatPhi}{\bimPhi}
\safemath{\rmatSigma}{\bimSigma}
\safemath{\rmatOmega}{\bimOmega}
\safemath{\rmatTheta}{\bimTheta}

\usepackage{amssymb}
\usepackage{amsfonts}
\usepackage{mathrsfs}
\usepackage{xspace}
\usepackage{bm}
\usepackage{fancyref}
\usepackage{textcomp}

\usepackage{multirow}
\usepackage{stmaryrd}

\newenvironment{textbmatrix}{	\setlength{\arraycolsep}{2.5pt}%
								\big[\begin{matrix}}{\end{matrix}\big]%
								\raisebox{0.08ex}{\vphantom{M}}}

\def\be{\begin{equation}}
\def\ee{\end{equation}}
\def\een{\nonumber \end{equation}}
\def\mat{\begin{bmatrix}}
\def\emat{\end{bmatrix}}
\def\btm{\begin{textbmatrix}}
\def\etm{\end{textbmatrix}}

\def\ba#1\ea{\begin{align}#1\end{align}}
\def\bas#1\eas{\begin{align*}#1\end{align*}}
\def\bs#1\es{\begin{split}#1\end{split}}
\def\bg#1\eg{\begin{gather}#1\end{gather}}
\def\bml#1\eml{\begin{multline}#1\end{multline}}
\def\bi#1\ei{\begin{itemize}#1\end{itemize}}

\newcommand{\lefto}{\mathopen{}\left}

\DeclareMathOperator{\Exop}{\opE}			
\DeclareMathOperator{\Varop}{\opV\!\mathrm{ar}} 

\newcommand{\Ex}[2]{\ensuremath{\Exop_{#1}\lefto[#2\right]}} 	
\newcommand{\abs}[1]{\lefto\lvert#1\right\rvert}		

\safemath{\dirac}{\delta}					
\safemath{\krond}{\dirac}					

\safemath{\upto}{\uparrow}
\safemath{\downto}{\downarrow}
\safemath{\iu}{j}							
\safemath{\ev}{\lambda}						
\safemath{\hilseqspace}{l^{2}}				
\newcommand{\banachfunspace}[1]{\setL^{#1}}	
\safemath{\hilfunspace}{\banachfunspace{2}}	

\safemath{\SNR}{\textit{SNR}} 				
\safemath{\PAR}{\textit{PAR}} 				
\safemath{\No}{N_0}							
\safemath{\Es}{E_s}							
\safemath{\Eb}{E_b}							
\safemath{\EbNo}{\frac{\Eb}{\No}}
\safemath{\EsNo}{\frac{\Es}{\No}}

\DeclareMathOperator{\CHop}{\ensuremath{\opH}} 
\safemath{\tvir}{\rndh_{\CHop}}				
\safemath{\tvtf}{\rndl_{\CHop}}				
\safemath{\spf}{\rnds_{\CHop}}				
\safemath{\bff}{H_{\CHop}}					

\safemath{\ircf}{r_{h}}						
\safemath{\tftvcf}{r_{s}}					
\safemath{\tfcf}{r_{l}}						
\safemath{\bfcf}{r_{H}}						

\safemath{\tcorr}{c_h}						
\safemath{\scf}{c_{s}}						
\safemath{\tfcorr}{c_{l}}					
\safemath{\fcorr}{c_{H}}						

\safemath{\mi}{I}							
\safemath{\capacity}{C}						

\safemath{\normal}{\mathcal{N}}			
\safemath{\jpg}{\mathcal{CN}}			
\safemath{\mchain}{\leftrightarrow}		

\safemath{\dB}{\,\mathrm{dB}}
\safemath{\dBm}{\,\mathrm{dBm}}
\safemath{\Hz}{\,\mathrm{Hz}}
\safemath{\kHz}{\,\mathrm{kHz}}
\safemath{\MHz}{\,\mathrm{MHz}}
\safemath{\GHz}{\,\mathrm{GHz}}
\safemath{\s}{\,\mathrm{s}}
\safemath{\ms}{\,\mathrm{ms}}
\safemath{\mus}{\,\mathrm{\text{\textmu}s}}
\safemath{\ns}{\,\mathrm{ns}}
\safemath{\ps}{\,\mathrm{ps}}
\safemath{\meter}{\,\mathrm{m}}
\safemath{\mm}{\,\mathrm{mm}}
\safemath{\cm}{\,\mathrm{cm}}
\safemath{\m}{\,\mathrm{m}}
\safemath{\W}{\,\mathrm{W}}
\safemath{\mW}{\, \mathrm{mW}}
\safemath{\J}{\,\mathrm{J}}
\safemath{\K}{\,\mathrm{K}}
\safemath{\bit}{\,\mathrm{bit}}
\safemath{\nat}{\,\mathrm{nat}}

\safemath{\define}{\triangleq}			

\safemath{\equivalent}{\sim}
\safemath{\distas}{\sim}					
\safemath{\sdiff}{\Delta}				

\safemath{\reals}{\mathbb{R}}
\safemath{\positivereals}{\reals_{+}}
\safemath{\integers}{\mathbb{Z}}
\safemath{\posint}{\integers_{+}}
\safemath{\naturals}{\mathbb{N}}
\safemath{\posnaturals}{\naturals_{+}}
\safemath{\complexset}{\mathbb{C}}
\safemath{\rationals}{\mathbb{Q}}

\newcommand*{\fancyrefapplabelprefix}{app}		
\newcommand*{\fancyrefthmlabelprefix}{thm}		
\newcommand*{\fancyreflemlabelprefix}{lem}		
\newcommand*{\fancyrefcorlabelprefix}{cor}		
\newcommand*{\fancyrefdeflabelprefix}{def}		
\newcommand*{\fancyrefproplabelprefix}{prop}		
\newcommand*{\fancyrefexmpllabelprefix}{exmpl}
\newcommand*{\fancyrefalglabelprefix}{alg}		
\newcommand*{\fancyreftbllabelprefix}{tbl}		

\frefformat{vario}{\fancyrefseclabelprefix}{Section~#1}
\frefformat{vario}{\fancyrefthmlabelprefix}{Theorem~#1}
\frefformat{vario}{\fancyreftbllabelprefix}{Table~#1}
\frefformat{vario}{\fancyreflemlabelprefix}{Lemma~#1}
\frefformat{vario}{\fancyrefcorlabelprefix}{Corollary~#1}
\frefformat{vario}{\fancyrefdeflabelprefix}{Definition~#1}
\frefformat{vario}{\fancyreffiglabelprefix}{Fig.~#1}
\frefformat{vario}{\fancyrefapplabelprefix}{Appendix~#1}
\frefformat{vario}{\fancyrefeqlabelprefix}{(#1)}
\frefformat{vario}{\fancyrefproplabelprefix}{Proposition~#1}
\frefformat{vario}{\fancyrefexmpllabelprefix}{Example~#1}
\frefformat{vario}{\fancyrefalglabelprefix}{Algorithm~#1}

 \newtheorem{thm}{Theorem}
 \newtheorem{cor}[thm]{Corollary}   
 
 \newtheorem{defi}{Definition}
 \newtheorem{lem}[thm]{Lemma}
 
\safemath{\dictab}{[\,\dicta\,\,\dictb\,]}

\safemath{\ysig}{\bmy}
\safemath{\ysighat}{\hat{\ysig}}
\safemath{\ysigdim}{M}
\safemath{\xsig}{\bmx}
\safemath{\xsigdim}{N}
\safemath{\nx}{n_x}
\safemath{\zsig}{\bmz}
\safemath{\zsigdim}{\ysigdim}
\safemath{\rsig}{\bmr}
\safemath{\Adict}{\bA}
\safemath{\Adicttilde}{\widetilde{\Adict}}
\safemath{\Adictdim}{\outputdim\times\xsigdim}
\safemath{\avec}{\bma}
\safemath{\avectilde}{\tilde{\avec}}
\safemath{\Bdict}{\bB}
\safemath{\Bdicttilde}{\widetilde{\Bdict}}
\safemath{\Cdict}{\bC}
\safemath{\cvec}{\bmc}
\safemath{\Ddict}{\bD}
\safemath{\Ddictdim}{\ysigdim\times\xsigdim}
\safemath{\dvec}{\bmd}
\safemath{\Ddicttilde}{\widetilde{\bD}}
\safemath{\Bonb}{\bB}
\safemath{\bvec}{\bmb}
\safemath{\Bonbdim}{\ysigdim\times\ysigdim}
\safemath{\noise}{\bmn}
\safemath{\noisedim}{\ysigim}
\safemath{\err}{\bme}
\safemath{\errdim}{\ysigdim}
\safemath{\errset}{\setE}
\safemath{\nerr}{n_e}
\safemath{\delop}{\bP_\errset}
\safemath{\delopc}{\bP_{{\errset}^c}}

\safemath{\cplxi}{\imath}
\safemath{\cplxj}{\jmath}

\safemath{\dict}{\matD}
\safemath{\inputdim}{N}		
\safemath{\outputdim}{M}		
\safemath{\sparsity}{S}	
\safemath{\inputdimA}{{N_a}}	
\safemath{\inputdimB}{{N_b}}	
\safemath{\elemA}{{n_a}}	
\safemath{\elemB}{{n_b}}	
\safemath{\resA}{\matR_a}	
\safemath{\resB}{\matR_b}	
\safemath{\subD}{\matS} 
\safemath{\subA}{\matS_a} 
\safemath{\subB}{\matS_b} 
\safemath{\dicta}{\matA} 	
\safemath{\dictb}{\matB} 	
\safemath{\hollowS}{H}
\safemath{\hollowA}{H_a}
\safemath{\hollowB}{H_b}
\safemath{\cross}{Z}
\safemath{\coh}{\mu_d}			
\safemath{\coha}{\mu_a}			
\safemath{\cohb}{\mu_b}			
\safemath{\mubs}{\nu}	
\safemath{\cohm}{\mu_m} 
\safemath{\dictset}{\setD}	
\safemath{\dictsetp}{\dictset(\coh,\coha,\cohb)}	
\safemath{\dictsetgen}{\dictset_\text{gen}}
\safemath{\dictsetgenp}{\dictsetgen(\coh)}
\safemath{\dictsetonb}{\dictset_\text{onb}}
\safemath{\dictsetonbp}{\dictsetonb(\coh)}

\safemath{\leftside}{U}
\safemath{\rightsideA}{R_a}
\safemath{\rightsideB}{R_b}

\safemath{\indexS}{\setI_S} 

\safemath{\na}{n_a}			
\safemath{\nb}{n_b}			
\safemath{\coeffa}{p_i}	
\safemath{\coeffb}{q_j}	
\safemath{\seta}{\setP}		
\safemath{\setb}{\setQ}     
\safemath{\setw}{\setW}	
\safemath{\setz}{\setZ}	
\safemath{\cola}{\veca}		
\safemath{\colb}{\vecb}		
\safemath{\cold}{\vecd}		
\safemath{\inputvec}{\vecx} 	
\safemath{\error}{\vece}	
\safemath{\noiseout}{\vecz} 	
\safemath{\inputvecel}{x}
\safemath{\inputveca}{\vecx_a}
\safemath{\inputvecb}{\vecx_b}
\safemath{\outputvec}{\vecy}	
\safemath{\lambdamin}{\lambda_{\mathrm{min}}}

\safemath{\elltwo}{\ell_2}
\safemath{\ellone}{\ell_1}
\safemath{\ellzero}{\ell_0}
\safemath{\ellinf}{\ell_\infty}
\safemath{\ellinftilde}{\ell_{\widetilde\infty}}
\safemath{\licard}{Z(\coh,\coha,\cohb)}
\safemath{\xsol}{\hat{x}}
\safemath{\xbord}{x_b}		
\safemath{\xstat}{x_s}		
\safemath{\xstatLone}{\tilde{x}_s}
\safemath{\order}{\mathcal{O}} 
\safemath{\scales}{\Theta} 
\safemath{\ones}{\mathbf{1}} 
\safemath{\zeroes}{\mathbf{0}} 
\safemath{\thlone}{\kappa(\coh,\cohb)} 
\safemath{\constoneA}{\delta} 
\safemath{\constoneB}{\epsilon} 
\safemath{\nlarge}{L}				   
\safemath{\sumlarge}{S_\nlarge}
\safemath{\maxlarger}{P_\nlarge}	   
\safemath{\Pzero}{\textrm{P0}}	
\safemath{\Pone}{\textrm{P1}}
\safemath{\vecfir}{\vecw}			 
\safemath{\vecsec}{\vecz}
\safemath{\elvecfir}{w}              
\safemath{\elvecsec}{z}				 
\safemath{\nlargefir}{n}
\safemath{\normout}{\gamma}
\safemath{\auxfun}{h}
\safemath{\supp}{\textrm{supp}}

\safemath{\indexa}{\ell}
\safemath{\indexb}{r}
\safemath{\indexc}{i}
\safemath{\indexd}{j}

\safemath{\project}{P}

\newcommand{\dd}{\textnormal{d}}%

\usepackage{color}

\safemath{\SINR}{\mathsf{sinr}}

\newtheorem{remark}{Remark}

\newcommand{\revision}[1]{#1}

\newcommand{\PD}{PD}

\newcommand{\FD}{FD}

\safemath{\LAMA}{\textrm{LAMA}}

\safemath{\mLAMA}{\textrm{M-LAMA}}
\safemath{\smLAMA}{\textrm{SM-LAMA}}
\safemath{\mCBAMP}{\textrm{mcB-AMP}}

\safemath{\MRT}{\textrm{MRT}}
\safemath{\betamax}{\beta^\textnormal{max}}
\safemath{\tmax}{t_\textnormal{max}}
\safemath{\betamaxno}{\beta^\textnormal{max}}
\safemath{\betamin}{\beta^\textnormal{min}}
\safemath{\betaminno}{\beta^\textnormal{min}}

\safemath{\Nomin}{\No^\textnormal{min}(\beta)}
\safemath{\Nominnobeta}{\No^\textnormal{min}}
\safemath{\Nomax}{\No^\textnormal{max}(\beta)}
\safemath{\Nomaxnobeta}{\No^\textnormal{max}}

\safemath{\MAP}{\textrm{MAP}}
\safemath{\IO}{\textrm{IO}}
\safemath{\JO}{\textrm{JO}}
\safemath{\Nopost}{N_{0}^\textnormal{post}}
\safemath{\MT}{U}
\safemath{\MR}{B}
\safemath{\C}{C}
\safemath{\Tran}{\textnormal{T}}
\safemath{\bmymf}{\bmy^\textnormal{MF}}
\safemath{\bmymrc}{\bmy^\textnormal{MRC}}
\safemath{\Herm}{\textnormal{H}}
\safemath{\row}{\textnormal{r}}
\safemath{\col}{\textnormal{c}}

\begin{document}

\title{Decentralized Equalization with Feedforward Architectures for Massive MU-MIMO}\author{Charles Jeon, Kaipeng Li, Joseph R. Cavallaro, and Christoph Studer\thanks{C.~Jeon was with the School of Electrical and Computer Engineering (ECE) at Cornell University, Ithaca, NY, and is now with Intel Labs, Hillsboro, OR; email: \url{cj339@cornell.edu}.}\thanks{C. Studer was with the School of ECE at Cornell University, Ithaca, NY, and is now with Cornell Tech, New York City, NY; email:  \url{studer@cornell.edu}; website: \url{vip.ece.cornell.edu}.}\thanks{K.~Li and J.~R.~Cavallaro are with the Department of ECE at Rice University, Houston, TX; email: \url{kl33@rice.edu}, \url{cavallar@rice.edu}.}
\thanks{Parts of the theoretical analysis in this paper have been presented at the 2017 IEEE International Symposium on Information Theory (ISIT)~\cite{jeon2017achievable}.}
}

\maketitle

\begin{abstract}
Linear data-detection algorithms that build on zero forcing~(ZF) or linear minimum mean-square error (L-MMSE) equalization achieve near-optimal spectral efficiency in massive multi-user multiple-input multiple-output (MU-MIMO) systems. 
Such algorithms, however, typically rely on centralized processing at the base-station~(BS) which results in (i) excessive interconnect and chip input/output (I/O) data rates and (ii) high computational complexity. 
Decentralized baseband processing (DBP) partitions the BS antenna array into independent clusters that are associated with separate radio-frequency circuits and computing fabrics in order to overcome the limitations of centralized processing.
In this paper, we investigate decentralized equalization with feedforward architectures that minimize the latency bottlenecks of existing DBP solutions.
We propose two distinct architectures with different interconnect and I/O bandwidth requirements that fuse the local equalization results of each cluster in a feedforward network.
For both architectures, we consider maximum ratio combining, ZF, L-MMSE, and a nonlinear equalization algorithm that relies on approximate message passing. \revision{For these algorithms and architectures, we analyze the associated post-equalization signal-to-noise-and-interference-ratio (SINR).}
We provide reference implementation results on a multi graphics processing unit (GPU) system which demonstrate that decentralized equalization with  feedforward architectures enables throughputs in the Gb/s regime and incurs no or only a small performance loss compared to centralized solutions. 
\end{abstract}


%


\section{Introduction}
\label{sec:intro}

\IEEEPARstart{M}{assive} multi-user (MU) multiple-input multiple-output (MIMO) will be a key technology for next-generation wireless systems~\cite{Marzetta10,mimo_overview,ABCHLAZ2014}.
By equipping the infrastructure base-stations (BSs) with hundreds or thousands of active antenna elements and serving tens or hundreds of user equipments (UEs) simultaneously and in the same frequency band, massive MU-MIMO promises orders-of-magnitude improvements in spectral efficiency and energy efficiency compared to traditional, small-scale MIMO \cite{HBD11,LETM2014}.
\revision{However, the large number of antennas at the BS causes significant challenges when implementing this technology in practice.}
One of the most prominent challenges is the excessively high amount of fronthaul data that must be transferred from the  radio-frequency (RF) antenna units at the BS antenna array to the baseband processing unit (BBU)~\cite{puglielli2015scalable,li2017decentralized,van2018efficient,jacobsson18c}.
For example, the fronthaul data rates (from RF chains to the BBU) exceed $200$\,Gbit/s for a massive MU-MIMO system with $128$ BS antennas, each using two $10$\,bit analog-to-digital converters (for in-phase and quadrature components) operating at $80$\,MS/s sampling rate.
Such high data rates not only exceed the bandwidth of existing high-speed interconnect standards, such as the common public radio interface (CPRI) \cite{cpri}, but will also approach the limits of existing chip input/output (I/O) interfaces in terms of bandwidth and power dissipation~\cite{puglielli2016design}.
\revision{Furthermore, traditional equalization-based data-detection algorithms that achieve near-optimal spectral efficiency in the MU-MIMO uplink~\cite{HBD11}, such as zero-forcing (ZF) and linear minimum mean-square error (L-MMSE)-based equalization, rely on centralized processing in a single computing fabric, which results in excessively high complexity and power consumption for systems with large antenna arrays~\cite{WYWDCS2014,li2017decentralized}.}

\subsection{Decentralized Baseband Processing}
In order to mitigate the bandwidth and computing bottlenecks of centralized massive MU-MIMO architectures, existing testbeds either distribute the most critical baseband processing tasks in the frequency domain or use maximum ratio combining (MRC).
\revision{Concretely, the testbeds described in~\cite{MVNKWOETL2016,lund,lund2017,YLYFTHZZ2013} parallelize the key baseband processing tasks across the subcarriers of orthogonal frequency-division multiplexing (OFDM)-based systems.} While this approach enables high parallelism, it requires that each frequency cluster obtains data from \emph{all} BS antennas, 
which alone does not enable one to scale such systems to thousands of antenna elements~\cite{li2017decentralized}.
In contrast to frequency parallelization, MRC enables antenna parallelization that divides array into independent clusters~\cite{Marzetta10,SYALMYZ2012}; this approach significantly reduces the interconnect bandwidth between the RF chains and the BBUs.
MRC, however, suffers from low spectral efficiency for realistic antenna configurations and high-rate modulation and coding schemes~\cite{HBD11}.
Consequently, realizing massive MU-MIMO in practice requires solutions that reduce the interconnect and chip I/O bandwidth as well as the baseband processing complexity per computing fabric, without sacrificing spectral efficiency.

Decentralized baseband processing (DBP) has been proposed in~\cite{li2017decentralized} to alleviate the fronthaul and I/O bandwidth bottlenecks, and enables parallel baseband processing across BS antennas on multiple computing fabrics, such as application-specific integrated circuits (ASICs), field-programmable gate arrays (FPGAs), or graphics processing units (GPUs)~\cite{LSCCGS2016,LCSGCS2016}, while achieving high spectral efficiency. 
The idea of DBP is to partition the BS antenna array into $C$ independent clusters, each associated with local computing fabrics that carry out the necessary RF and baseband processing tasks in a decentralized and parallel fashion. 
The algorithms proposed in~\cite{li2017decentralized} perform linear equalization and precoding in an iterative manner by exchanging consensus information among the clusters. 
However, implementation results on a GPU cluster revealed that the transfer latency of such consensus-sharing methods are limiting the achievable throughput. 
To avoid this drawback, references~\cite{puglielli2015scalable,bertilsson2016scalable,puglielli2016design,jeon2017achievable}  recently proposed \emph{feedforward} architectures that minimize the transfer latency.

\subsection{Contributions}
We propose two distinct {feedforward} architectures for partially decentralized (PD) and fully decentralized (FD) equalization, which mitigate the interconnect, I/O, latency, and computation bottlenecks.
For both of these architectures, we investigate the efficacy of MRC, ZF, L-MMSE, and a nonlinear equalization method that builds upon the large-MIMO approximate message passing (LAMA) algorithm~\cite{JGMS2015conf}.
Our main contributions can be summarized as follows:

\begin{itemize}
\item We develop a framework that enables a precise analysis of the post-equalization signal-to-noise-and-interference-ratio (SINR) of decentralized equalization with feedforward architectures in the large-system limit.
\item We show that the PD feedforward architecture achieves the same SINR performance as centralized solutions for equalization with MRC, ZF, L-MMSE, and PD-LAMA.
\item We show that the FD feedforward architecture is able to provide near-optimal SINR performance, but further reduces the interconnect and I/O bandwidths. 
\item We analyze optimal antenna partitioning strategies that maximize the SINR for the FD architecture.
\item We conduct error-rate simulations for a realistic 3GPP long-term evolution (LTE)-like massive MU-MIMO system that support our theoretical findings. 
\item We provide reference throughput and latency results for linear and nonlinear equalization in centralized, PD, and FD architectures on a multi-GPU system.
\end{itemize}
Our results demonstrate that feedforward equalization enables throughputs in the Gb/s regime for massive MU-MIMO systems with hundreds of antenna elements, and incurs no or only a small loss in post-equalization SINR and error-rate performance compared to that of centralized solutions.

\subsection{Relevant Prior Art}

DBP for massive MI-MIMO systems has been proposed in~\cite{li2017decentralized} together with consensus-sharing equalization and precoding algorithms. 
Distributed processing across antenna elements is also a critical component of coordinated multipoint (CoMP)~\cite{IDMGFBMTJ2011} and cloud radio access networks (CRANs) \cite{PLZW2015} for multi-cell transmission.  
While all these architectures and algorithms are able to reduce the raw baseband data rates and mitigate the computation bottlenecks, their performance has not been analyzed and the achievable throughput suffers from high interconnect latency caused by iterative exchange of consensus information.
\revision{To avoid iterative consensus sharing among antenna clusters, we focus on decentralized \emph{feedforward} architectures that minimize the transfer latency and enable a precise theoretical performance analysis.}

Feedforward architectures for decentralized massive MU-MIMO equalization have been proposed in~\cite{puglielli2015scalable,bertilsson2016scalable,puglielli2016design,jeon2017achievable}. 
The present paper extends our theoretical results from~\cite{jeon2017achievable} and, in contrast to~\cite{puglielli2015scalable,bertilsson2016scalable,puglielli2016design}, provides two distinct architectures and a corresponding SINR analysis for a range of linear  and nonlinear equalization algorithms. 
In addition, we provide reference implementation results on a GPU cluster to assess the throughput and latency of our architectures and algorithms. 

The post-equalization SINR performance of \emph{centralized} linear equalization algorithms, such as MRC, ZF, and L-MMSE, has been analyzed in \cite{VS1999,TH1999,SV2001,ghods2017optimally} in the large-system limit.
We will investigate the SINR performance of these algorithms for the two proposed decentralized feedforward architectures, and also investigate the efficacy of nonlinear equalization for decentralized massive MU-MIMO architectures.

Nonlinear equalization for massive MU-MIMO systems via approximate message passing (AMP) has been studied in \cite{WKNLHG14,JGMS2015conf,JMS2016}. \revision{Message passing has also been used recently for data detection in non-orthogonal multiple access (NOMA)  systems~\cite{LYGLH2019}}.
A distributed version of AMP has been proposed in~\cite{ZBB2016} for compressive sensing applications \cite{donoho2006,CRT06}.
\revision{The key differences of our nonlinear equalization algorithm to these results are as follows: (i) We consider decentralized feedforward architectures; (ii) the methods in~\cite{WKNLHG14,JGMS2015conf,JMS2016,LYGLH2019} are centralized; (iii) the distributed AMP-based method in~\cite{ZBB2016} requires iterative consensus sharing; (iv) we analyze the post-equalization SINR and error-rate performance in massive MU-MIMO systems.}

\subsection{Notation}
 Lowercase and uppercase boldface letters designate vectors and matrices, respectively; uppercase calligraphic letters denote sets. The transpose and Hermitian of the matrix $\bA$ are represented by $\bA^\Tran$ and $\bA^\Herm$, respectively. The $M\times N$ all-zeros matrix is~$\bZero_{M\times N}$ and the \revision{$N$-dimensional} identity matrix is $\bI_N$. The $k$th entry of a vector $\bma$ is $a_k$.
We define $\left\langle \bmx \right\rangle = \frac{1}{N}\sum_{k=1}^N x_k$. 
The circularly-symmetric multivariate complex-valued Gaussian probability density function (pdf) with covariance~$\bK$ is denoted by $\setC\setN(\bZero,\bK)$. $\Exop_X\!\left[\cdot\right]$ and $\Varop_X\!\left[\cdot\right]$ represent the mean and variance with respect to the 
random variable~$X$, respectively.

\subsection{Paper Outline}
The rest of the paper is organized as follows. 
\fref{sec:Decentarch} introduces the system model and the two feedforward architectures. 
\fref{sec:PDequalization} and \fref{sec:FDequalization} investigate equalization algorithms for the PD and FD architecture, respectively. 
\fref{sec:Results} provides theoretical and simulative results for the proposed methods.
\fref{sec:gpuimplementation} details the multi-GPU implementation and provides throughput and latency results.
\fref{sec:conclusions} concludes the paper.
All derivations and proofs are relegated to the appendices.


\begin{figure*}[tp]
\centering
\subfigure[Partially decentralized (PD) equalization architecture.]{\includegraphics[height=0.45\columnwidth]{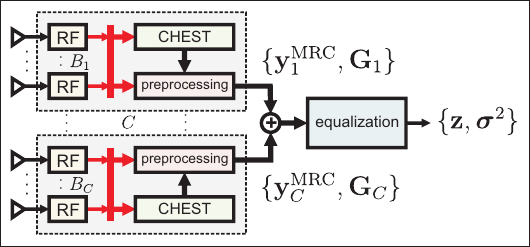}
\label{fig:pda}
}
\hspace{0.9cm}
\subfigure[Fully decentralized (FD) equalization architecture.]{\includegraphics[height=0.45\columnwidth]{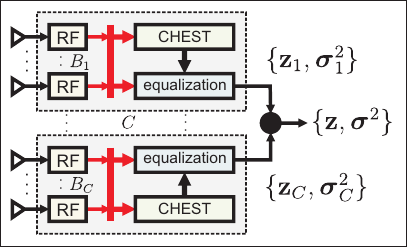}
\label{fig:fda}
}
\caption{Partially decentralized (PD) and fully decentralized (FD) feedforward equalization architectures for the massive MU-MIMO uplink. The antenna array is divided in $\C$ clusters, each associated with local radio-frequency (RF) processing and channel estimation (CHEST).
(a) PD performs decentralized CHEST and preprocessing; equalization is performed in a centralized fashion and operates on a low-dimensional data (dimension is the number of UEs). (b) FD performs CHEST, preprocessing, and equalization in a decentralized manner; the final equalization result is formed by a weighted average of local estimates. 
The $\oplus$ operator in (a) denotes matrix/vector-additions and $\bullet$ in (b) denotes a weighted vector addition; see Eq.~\ref{eq:opt_fusion} in \fref{sec:FDequalization} for the details.
}
\label{fig:architectures}
\end{figure*}

\section{Decentralized Equalization Architectures}\label{sec:Decentarch}
We start by introducing the considered massive MU-MIMO system model and the basics of equalization-based data detection. We  then discuss the two feedforward equalization architectures for DBP depicted in \fref{fig:architectures}, and detail the SINR analysis framework that we will use throughout the paper.

\subsection{Uplink System Model and Equalization}
We consider a narrowband massive MU-MIMO uplink system in which $U$ single-antenna UEs transmit data to a BS with $\MR$ antenna elements. 
To model this scenario, we use the standard input-output relation~\cite{Marzetta10}
\begin{align} \label{eq:iorelation}
\bmy=\bH\bms_0+\bmn.
\end{align} 
Here, $\bmy\in\complexset^\MR$ is the receive vector at the BS, $\bH\in\complexset^{\MR\times\MT}$ represents the MIMO system matrix, which we assume is perfectly known at the BS,  $\bms_0\in\setO^\MT$ contains the transmit symbols for each UE, $\setO$ is the constellation set (e.g., QPSK or 16-QAM), and  $\bmn\in\complexset^\MR$ is i.i.d.\ circularly symmetric complex Gaussian noise with variance $\No$ per complex entry.
We assume an i.i.d. prior $p(\bms_0)=\prod_{u=1}^{\MT} p(s_{0u})$ for the transmit vector  and the following distribution for each transmit symbol:
\begin{align}\label{eq:prior}
p(s_{0u}) =   \frac{1}{|\setO|} \sum_{a\in\setO}  \delta(s_{0u} -a),
\end{align}
where $|\setO|$ is the cardinality of the constellation $\setO$  and  $\delta(\cdot)$ is the Dirac delta function. In what follows, we assume zero-mean constellations and define the average energy per transmit symbol as $\Es=\Ex{}{|s_{0u}|^2}$, $u=1,\ldots,U$.

Equalization is concerned with forming an estimate $\bmz$ of the transmit signal vector $\bms_0$ along with reliability estimates for each entry in $\bmz$. \revision{These two quantities are then used by the data detector to compute hard-output estimates for the transmit symbols or bit-wise soft information in the form of log-likelihood ratios~\cite{paulraj03,studer2011asic}. }
Consider a general \emph{centralized} equalizer $\{\bmz,\boldsymbol\sigma^2\} = \setE(\bmy,\bH)$ that takes the received vector~$\bmy$ and the MIMO channel matrix~$\bH$ in order to compute (i) an estimate~$\bmz$ for the true transmit vector~$\bms_0$ and (ii) the associated error variance vector $\boldsymbol\sigma^2$.
The error variance vector characterizes the post-equalization residual interference and noise variance on each entry of the estimate~$\bmz$. Mathematically, this quantity corresponds to the variances of each entry in the residual interference and noise vector defined as
$\bme=\bmz-\bms_0$, i.e.,  
$\boldsymbol\sigma^2 = \Ex{}{|\bme|^2}$ where $|\cdot|^2$ operates element-wise on vectors. 

The literature describes a range of linear and nonlinear equalization algorithms for small-scale and massive MU-MIMO data detection \cite{WYWDCS2014,wu2016efficient,JGMS2015conf,pan2014mimo}. Linear methods, such as MRC, ZF, and L-MMSE are among the most common algorithms, mainly due to their simplicity and low computational complexity \cite{WYWDCS2014}. \revision{Nevertheless, nonlinear equalizers, such as the LAMA algorithm put forward in \cite{JGMS2015conf}, have been shown to (often significantly) outperform linear equalizers at the cost of higher computational complexity \cite{GV2005,JGMS2015conf,JMS2016}.}

\subsection{Basics of Decentralized Equalization}

As in \cite{li2017decentralized}, we partition the $\MR$ BS antenna elements into $C\in\{1,2,\ldots,B\}$ independent \emph{antenna clusters}. The $c$th antenna cluster is associated with $\MR_c=w_c\MR$ BS antennas so that  $w_c\in[0,1]$ and $\sum_{c=1}^{\C}w_c =1$.
Each cluster contains local RF components and only requires access to local channel state information~(CSI) acquired in a local channel estimation (CHEST) unit. 
Without loss of generality, we partition the receive vector $\bmy=[\bmy_1^\Tran,\ldots,\bmy_\C^\Tran]^\Tran$, the 
channel matrix $\bH = [\bH_1^\Tran,\ldots,\bH_\C^\Tran]^\Tran$, and the noise vector $\bmn=[\bmn_1^\Tran,\ldots,\bmn_\C^\Tran]^\Tran$ in~\fref{eq:iorelation}.
For this antenna partitioning scheme, the input-output relation corresponding to the local receive vector~$\bmy_c$ associated with the $c$th cluster can be written 
as
\begin{align} \label{eq:perclusterInOutrelation}
\bmy_c=\bH_c\bms_{0}+\bmn_c, \quad c=1,\ldots,C,
\end{align} 
with $\bmy_c\in\complexset^{\MR_c}$, $\bH_c\in\complexset^{\MR_c\times\MT}$, and $\bmn_c\in\complexset^{\MR_c}$. 
The following subsections describe two decentralized equalization architectures that compute estimates for the transmit vector~$\bms_{0}$ by performing local computations in each antenna cluster using only information of the local receive vector $\bmy_c$ and channel matrix $\bH_c$ followed by fusion of the  results from all clusters.
\subsection{Partially Decentralized (\PD) Equalization Architecture}\label{sec:Decentarch_CE}
The partially decentralized (\PD) equalization architecture is illustrated in \fref{fig:pda}.
First, each cluster~$c=1,\ldots,C$ independently (and in parallel) preprocesses the local receive vector~$\bmy_c$ and channel matrix~$\bH_c$ by computing the $U$-dimensional local MRC vector  $\bmymrc_c = \bH_c^\Herm\bmy_c$ and the $U\times U$ local Gram matrix $\bG_c=\bH_c^\Herm \bH_c$.
Second, a \emph{feedforward} adder tree, indicated with the symbol $\oplus$  in \fref{fig:pda}, is used to compute the complete MRC vector and Gram matrix as follows: 
\begin{align} \label{eq:fullMRC}
\bmymrc=\sum_{c=1}^C\bmymrc_c \quad \text{and} \quad \bG=\sum_{c=1}^\C \bG_c.
\end{align}
Third, we perform linear or nonlinear equalization in a centralized unit that computes the 
estimate $\bmz\in\complexset^\MT$ and the post-equalization error variance vector $\boldsymbol\sigma^2\in\complexset^\MT$. The tuple $\{\bmz,\boldsymbol\sigma^2\}$ is then used to compute hard- or soft-output estimates.

In \fref{sec:PDequalization}, we will detail MRC, ZF, L-MMSE equalization, and a new LAMA-based equalization algorithm~\cite{JGMS2015conf} for the PD architecture, all of which directly operate on the $U$-dimensional fused MRC vector $\bmymrc$ and Gram matrix $\bG$. 
Since the MRC vector is a sufficient statistic for the transmit signal $\bms_0$~\cite{paulraj03}, we will show that the PD equalization does not incur an SINR performance loss compared to centralized MRC, ZF, L-MMSE, and LAMA equalizers.

\subsection{Fully Decentralized (\FD) Equalization Architecture}\label{sec:decent_fda}
The \PD{} architecture requires a summation of both the local MRC vectors and the local Gram matrices, which involves potentially large amounts of data to be transmitted to the central equalization unit, especially in channels with short coherence time.
The fully decentralized (\FD) equalization architecture illustrated in \fref{fig:fda} avoids the transmission of the local Gram matrices altogether at the cost of a (typically small) performance loss.
First, each cluster $c=1,\ldots,C$ independently (and in parallel) performs CHEST, preprocessing, \emph{and} equalization, i.e., directly
forms a local estimate $\bmz_c\in\complexset^\MT$ and local post-equalization error variance vector $\boldsymbol\sigma_c^2\in\complexset^\MT$.
Second, a feedforward fusion tree, indicated with the symbol~$\bullet$  in \fref{fig:fda},  optimally combines the local estimates~$\bmz_c$ using information from the error variance vectors $\boldsymbol\sigma_c^2$ in order to generate the final output tuple $\{\bmz,\boldsymbol\sigma^2\}$.

In \fref{sec:FDequalization}, we will detail the optimal fusion rule as well as MRC, ZF, L-MMSE, and LAMA-based equalization for the FD architecture.  We will also provide an SINR performance analysis in the large-system limit.

\subsection{Signal-to-Interference-and-Noise-Ratio (SINR) Analysis}
\label{sec:SINRanalysis}
To analyze the performance of linear and nonlinear equalization algorithms for the PD and FD architectures, we will focus on the large-system limit and  Rayleigh-fading channels. Hence, we will make frequent use of the following two definitions. 

\begin{defi}[Large-system limit] The \emph{large-system limit} is defined by fixing the \emph{system ratio} $\beta=\MT/\MR$ and $\MT,\MR\to\infty$. 
\end{defi}

\begin{defi}[Rayleigh fading] A MIMO channel is \emph{Rayleigh fading} if the channel matrix $\bH$ has i.i.d.\ circularly symmetric complex Gaussian entries with variance $1/\MR$ per entry.
\end{defi}

By considering the large-system limit and Rayleigh-fading channels, Tse and Hanly have shown in~\cite{TH1999} that linear equalizers, such as MRC, ZF, and L-MMSE, \emph{decouple} MIMO systems into parallel and independent additive white Gaussian noise (AWGN) channels. This means that the estimate $\bmz$ of such linear equalizers can be modeled on a per-UE basis in a statistically equivalent manner as follows: 
\begin{align} \label{eq:decoupling}
z_u = s_{0u} + e_u, \quad u=1,\ldots,U,
\end{align}
where $e_u\in\complexset$ represents residual interference and noise. Furthermore, the quantity $e_u$ turns out to be (i) statistically independent of $s_{0u}$ and (ii) circularly symmetric complex Gaussian with \emph{decoupled noise variance} $\sigma^2$, which does not depend on the UE index~$u$. 
This result also implies that all entries of the error variance vector $\boldsymbol\sigma^2$ correspond to $\sigma^2$. 
In \fref{sec:PDequalization} and \fref{sec:FDequalization} for the PD and FD architecture, respectively, we will  build upon this asymptotic analysis framework in order to theoretically characterize the per-UE post-equalization SINR of the decoupled system~\fref{eq:decoupling}:
\begin{align} \label{eq:SINRdefinition}
\SINR \define \frac{\Es}{\sigma^2}.
\end{align} 
Numerical results that validate our asymptotic analysis in finite-dimensional systems will be presented in \fref{sec:Results}.


\section{Partially Decentralized (\PD) Equalization} \label{sec:PDequalization}
We start by reviewing linear equalization algorithms for the PD architecture depicted in \fref{fig:pda}, and adapt the well-known Tse-Hanly equations~\cite{TH1999} to analyze the associated post-equalization SINR performance in the large-system limit. 
We then present a new, nonlinear equalization algorithm that builds upon LAMA proposed in~\cite{JGMS2015conf}, and we develop a corresponding SINR performance analysis for the PD architecture. 

\subsection{Linear Equalization Algorithms for the \PD{} Architecture}
Since the MRC output $\bmymrc$ in \fref{eq:fullMRC} is a sufficient statistic for the transmit signal vector $\bms_0$, a variety of optimal and suboptimal equalization-based data detection algorithms can be derived from this quantity~\cite{paulraj03}. 
For MRC-based equalization in the PD architecture, we use~\fref{eq:fullMRC} to form the estimate 
\begin{align*}
\bmz^\text{MRC} = \mathrm{diag}(\bG)^{-1}\bmymrc,
\end{align*} 
where the diagonal matrix $\mathrm{diag}(\bG)^{-1}$ is computed in the centralized equalization unit; see~\fref{fig:pda}.
The MRC estimate~$\bmz^\text{MRC}$ can then be used to perform either hard- or soft-output data detection. 
For soft-output data detection, one requires the error variance vector given by
\begin{align*}
\boldsymbol\sigma_\text{MRC}^2 =\,\,& \mathrm{diag}  \big(\mathrm{diag}(\bG)^{-1}\bG \mathrm{diag}(\bG)^{-\Herm}\No  \\
& +  \left(\mathrm{diag}(\bG)^{-1}\bG-\bI_U\right)\!\left(\mathrm{diag}(\bG)^{-1}\bG-\bI_U\right)^\Herm\!\Es   \big)
\end{align*}
that contains the post-equalization SINR for each entry of~$\bmz^\text{MRC}$.
Note that MRC-based equalization was shown to be optimal (i) for a fixed number of UEs and infinitely many BS antennas~\cite{Marzetta10}, which is equivalent to $\beta\to0$ in the large-system limit, or (ii) in the low-SNR regime~\cite{TH1999}. 
The estimate of the ZF equalizer for the PD architecture is given by 
\begin{align*}
\bmz^\text{ZF} = \bG^{-1}\bmymrc,
\end{align*} 
where the matrix  $ \bG^{-1}$ is computed in the centralized equalization unit. The associated error variance vector is given by
\begin{align*}
\boldsymbol\sigma_\text{ZF}^2  =  \mathrm{diag}(\bG^{-1})\No.
\end{align*}
For L-MMSE equalization, we have
\begin{align*}
\bmz^\text{L-MMSE} = (\bG+\rho\bI_U)^{-1}\bmymrc.
\end{align*} 
where the matrix $(\bG+\rho\bI_U)^{-1}$ is computed in the centralized equalization unit. The L-MMSE regularization parameter is set to $\rho=\No/\Es$ for complex-valued constellations (e.g., QPSK or 16-QAM). 
The associated error variance vector is given by 
\begin{align*}
& \boldsymbol\sigma_\text{L-MMSE}^2  = \mathrm{diag} \big( (\bG+\rho\bI_U)^{-1}\bG(\bG+\rho\bI_U)^{-\Herm}\No  \\
& \quad + \left((\bG+\rho\bI_U)^{-1}\bG-\bI_U\right)\!\left((\bG+\rho\bI_U)^{-1}\bG-\bI_U\right)^\Herm\!\Es  \big)\!.
\end{align*}
We reiterate that the MRC, ZF, and L-MMSE equalizers for the PD architecture deliver exactly the \emph{same estimates} as their centralized counterparts---the only difference is the way the involved quantities are computed. 

As shown in~\cite{TH1999} and outlined in \fref{sec:SINRanalysis}, centralized MRC, ZF, and L-MMSE equalizers decouple MIMO systems in the large-system limit and for Rayleigh fading channels; this implies that the entries of the error variance vectors~$\boldsymbol\sigma_\text{MRC}^2$, $\boldsymbol\sigma_\text{ZF}^2$, and $\boldsymbol\sigma_\text{L-MMSE}^2$ converge to the decoupled noise variance~$\sigma^2$ of the MRC, ZF, and L-MMSE equalizer, respectively.
Since linear equalizers in the PD architecture yield exactly the same estimates as in a centralized architecture, we can directly characterize the associated decoupled noise variance~$\sigma_\text{PD}^2$ in the PD architecture using the following result.
\begin{thm}[\kern-0.33em\mbox{\cite[Thm.~3.1]{TH1999}}] \label{thm:THeq}
Fix the system ratio $\beta=\MT/\MR$, and assume the large-system limit and Rayleigh fading channels.
Then, the decoupled noise variance $\sigma_{\text{PD}}^2$ for MRC, ZF, and L-MMSE equalization in a centralized or PD architecture, is the solution to the following fixed-point equation
\begin{align} \label{eq:fixedpointequation}
\sigma_{\text{PD}}^2 = \No + \beta\Psi(\sigma_{\text{PD}}^2),
\end{align}
where the MSE function $\Psi(\sigma^2)$ is given by 
\begin{align*}
\Psi(\sigma^2) &= \Es, \tag{MRC}\\
\Psi(\sigma^2) &= \sigma^2,  \text{ for } \beta<1,\tag{ZF} \\
\Psi(\sigma^2) & = \frac{\Es}{\Es+\sigma^2}\sigma^2,  \tag{L-MMSE}
\end{align*}
for  MRC, ZF, and L-MMSE equalization, respectively.
\end{thm}
\revision{We note that the expression for the ZF equalizer only holds for \mbox{$\beta<1$}, whereas the expressions for MRC and L-MMSE hold for general system ratios~$\beta\geq0$.\footnote{The asymptotic SINR performance of ZF equalization via the Moore-Penrose pseudo inverse when~$\beta \geq1$ was analyzed in \cite{EC2003}.}}
From \fref{thm:THeq}, we obtain closed-form expressions for the post-equalization \SINR in \fref{eq:SINRdefinition} for MRC, ZF, and L-MMSE equalization in the PD architecture. 

\begin{cor}\label{cor:sinr_PD}
Assume that the conditions of \fref{thm:THeq} hold. Then, the post-equalization \SINR for MRC, ZF, and L-MMSE equalization in the PD architecture are given by 
\begin{align*}
\SINR^\textnormal{MRC}_\textnormal{PD} = \,& \frac{\Es/\No}{1+\beta\Es/\No}, \tag{MRC}\\
\SINR^\textnormal{ZF}_\textnormal{PD} = \,&  \frac{\Es}{\No}(1-\beta), \text{ for } \beta<1, \tag{ZF}\\
\SINR^\textnormal{L-MMSE}_\textnormal{PD} = \,&  
\frac{1}{2}
\Bigg(
\sqrt{
\left(1 - \frac{\Es}{\No}(1-\beta)\right)^{\!\!2} + 4\frac{\Es}{\No}}
\\
&- 
\Big(1 - \frac{\Es}{\No}(1-\beta)\Big)\!
\Bigg). \tag{L-MMSE}
\end{align*}
\end{cor}

We note that in the massive MU-MIMO regime, which corresponds to $\beta\to0$, all post-equalization SINR expressions converge to $\Es/\No$, which confirms the well-known fact that MRC is optimal in this regime~\cite{Marzetta10}.
It can also be shown that $\SINR^\textnormal{L-MMSE}_\textnormal{PD}$ bounds $\SINR^\textnormal{MRC}_\textnormal{PD}$ and $\SINR^\textnormal{ZF}_\textnormal{PD}$ from above for all system ratios and in all SNR regimes. Hence, L-MMSE equalization is often the preferred choice in realistic massive MU-MIMO systems~\cite{HBD11,WYWDCS2014}.
We reiterate that the SINR expressions listed in \fref{cor:sinr_PD} are also valid for centralized architectures. 

\subsection{LAMA for the \PD{} Architecture}\label{sec:LAMA_alg}
The LAMA algorithm developed in \cite{JGMS2015conf} is a nonlinear equalizer is able to achieve individually-optimal performance in the large-system limit given certain conditions on the antenna ratio $\beta$ and the noise variance $\No$ are satisfied. LAMA operates directly on the input-output relation in \fref{eq:iorelation} and is, hence, designed for centralized processing.
We now develop a novel variant of LAMA that directly operates on the complete MRC output~$\bmymrc$ and the Gram matrix $\bG$ in \fref{eq:fullMRC} to enables its use in the PD architecture.
Since the antenna configuration in massive MU-MIMO systems typically satisfies~\mbox{$\MT\ll\MR$}, the  LAMA-\PD{} algorithm  operates on a lower dimension which reduces complexity while delivering exactly the same estimates as the original LAMA algorithm. 
We note that LAMA was derived in the large-system limit and for Rayleigh fading channels~\cite{DMM10a}, but these assumptions are not required in practice.  
We next summarize the LAMA-PD algorithm; the derivation can be found in \fref{app:LAMAPD_derivation}.

\newtheorem{alg}{Algorithm}
\begin{alg}[LAMA-PD] Initialize $s_\ell=\Exop_S[S]$ for $\ell=1,\ldots,\MT$,  $\phi^{(1)} = \Varop_S[S]$, and $\bmv^{(1)}=\mathbf{0}_{\MR\times0}$. Then, for every  iteration $t=1,2,\ldots,T_\text{max}$, compute the following steps:\label{alg:lamanew}
\begin{align} \label{eq:LAMA_Gaussian}
\bmz^t &= \bmymrc + (\bI - \bG)\bms^t + \bmv^t \\
\nonumber\bms^{t+1} &= \mathsf{F}(\bmz^t, \No+\beta\phi^t)\\
\nonumber\phi^{t+1} &= \langle\mathsf{G}(\bmz^t, \No+\beta\phi^t)\rangle\\
\nonumber\bmv^{t+1} &=  \frac{\beta\phi^{t+1}}{\No+\beta\phi^t }(\bmz^t-\bms^t).
\end{align}
The functions $\mathsf{F}(s_\ell,\tau)$ and $\mathsf{G}(s_\ell,\tau)$ are the message mean and variance, respectively, operate element-wise on vectors, and are  computed as follows:
\begin{align}\label{eq:F}
\mathsf{F}(z_\ell,\tau) &=  
\int_{s_\ell} s_\ell f(s_\ell \vert \hat{z}_\ell) \dd s_\ell\\\nonumber
\mathsf{G}(z_\ell,\tau) &= \int_{s_\ell} \abs{s_\ell}^2 f(s_\ell \vert \hat{z}_\ell) \dd s_\ell - \abs{\mathsf{F}(z_\ell,\tau)}^2\!.
\end{align}
Here, $f(s\vert {z})$ is the posterior pdf $f(s\vert  z) = \frac{1}{Z}p( z \vert s)p(s)$ with $p( z\vert s)\sim \setC\setN(s,\tau)$, $p(s)$ is given in~\fref{eq:prior}, and $Z$ is a normalization constant. 
The estimates and error variances of LAMA are $\bmz^t$ and $\sigma^2_{t,\textnormal{LAMA}} = \No+\beta\phi^t$, respectively. 
\label{alg:LAMA_alg}
\end{alg}

In order to analyze the post-equalization SINR of LAMA-\PD{}, it is key to realize that the equalization output~$\bmz^{t}$ is equivalent to that of the original centralized LAMA algorithm in \cite{JGMS2015conf}. 
As shown in \cite{JGMS2015conf}, LAMA (and hence LAMA-PD) decouples the MIMO system into parallel AWGN channels.
More specifically, the equalizer output \fref{eq:LAMA_Gaussian} of LAMA can be modeled as in \fref{eq:decoupling}, where the decoupled noise variance $\sigma_{t}^2$ at iteration $t$ can be tracked using the state evolution (SE) framework in the large-system limit and for Rayleigh fading channels.
The following result, with proof in~\cite{BM2011}, summarizes this SE framework. 

\begin{thm}[\kern-0.33em\mbox{\cite[Thm. 1]{JGMS2015conf}}] 
Fix the system ratio $\beta=\MT/\MR$ and the signal prior \fref{eq:prior}. Assume the large-system limit and Rayleigh fading channels. 
Then, the decoupled noise variance~$\sigma_t^2$ of LAMA and LAMA-PD at iteration $t$ is given by the recursion: \label{thm:SE}
\begin{align}\label{eq:SE_recursion}
\sigma_t^2 = \No + \beta\Psi(\sigma_{t-1}^2).
\end{align}
Here, the MSE function is given by
\begin{align}\label{eq:SE_MSEpsi}
\Psi(\sigma_{t-1}^2)=\Exop_{S,Z}
\!\Big[
\abs{
  \mathsf{F}(S+\sigma_{t-1}Z,\sigma_{t-1}^2) - S
}^2
\Big],
\end{align}
where the function $\mathsf{F}$ is given in \fref{eq:F}, $S\sim p(s)$ as in \fref{eq:prior}, $Z\sim\setC\setN(0,1)$, and $\sigma_1^2$ is initialized by $\sigma_1^2=\No+\beta\Es$.
\end{thm}

For $t\to\infty$, the SE recursion in \fref{eq:SE_recursion} 
converges to the same fixed-point equation of linear equalizers in \fref{eq:fixedpointequation}, where the only difference is the MSE function \fref{eq:SE_MSEpsi}.
If there are multiple fixed points, then we select the largest $\sigma_{\text{PD}}^2$, which is, in general, a sub-optimal solution.\footnote{See \cite{GV2005} for more details on the existence of multiple fixed-points and on conditions for which LAMA achieves individually optimal performance.}
As for linear equalizers, we can use the fixed-point equation in  \fref{eq:fixedpointequation} to analyze the post-equalization SINR performance of LAMA and LAMA-PD.
Unfortunately, there are no closed-form expressions known for the decoupled noise variance or the SINR  for LAMA and LAMA-PD with discrete constellations, due to the specific form of the MSE function~\fref{eq:SE_MSEpsi}.
Nevertheless, we can numerically compute~\fref{eq:SE_MSEpsi} and, hence, analyze the SINR.
A corresponding SINR comparison with linear equalizers is given in \fref{sec:Results}.

\section{Fully Decentralized (\FD) Equalization}
\label{sec:FDequalization}

We next discuss optimal fusion for linear and nonlinear equalization in the  PD architecture as depicted in \fref{fig:fda}. We then analyze the post-equalization SINR of the proposed equalizers depending on the antenna cluster allocation strategy.

\subsection{Optimal Fusion for the \FD{} Architecture}\label{sec:FDA_algo}
As detailed in \fref{sec:decent_fda}, each cluster $c=1,\ldots,C$ in the \FD{} architecture independently computes a local estimate~$\bmz_c$ and associated error variance vector~$\boldsymbol\sigma_c^2$.
Then, the vectors~$\bmz_c$ and~$\boldsymbol\sigma_c^2$ for $c=1,\ldots,C$ are fused  to compute the final output tuple~$\{\bmz,\boldsymbol\sigma^2\}$.
Since in the large-system limit and for Rayleigh fading channels, the considered equalizers decouple MIMO systems into parallel and independent AWGN channels (see \fref{sec:SINRanalysis}), we focus on \emph{linear} fusion of the local estimates, indicated with $\bullet$  in \fref{fig:fda}. Specifically, the proposed FD architecture computes the fused estimate $\bmz$ by combining the local estimates for each UE as follows:
\begin{align} \label{eq:fusionrule}
z_u = \sum_{c=1}^C\nu_{c,u} z_{c,u}, \quad u=1,\ldots,U.
\end{align}
Here, $z_{c,u}$ is the local estimate for UE $u$ at cluster $c$ and the weights $\nu_{c,u}$, $u=1,\ldots,U$, depend on the per-cluster error variance vector~$\boldsymbol\sigma_c^2$. 
In what follows, we are interested in the optimal set of weights for the following criterion.
\begin{defi}[Optimal fusion] \label{def:optfusion}
Optimal fusion for the FD architecture maximizes the per-UE  post-equalization SINR of the final estimate $z_{u}$ obtained from \fref{eq:fusionrule} while $\sum_{c=1}^C\nu_{c,u}=1$.
\end{defi}
In other words, optimal fusion defines a set of weights $\nu_{c,u}$, $c=1,\ldots,C$, $u=1,\ldots,U$, so that the decoupled noise variances contained in~$\boldsymbol\sigma^2$ associated with the fused estimate~$\bmz$ are minimized. 
The following result summarizes the optimal fusion rule; a short proof is given in \fref{app:optimal_fusion}.
\begin{lem}
\label{lem:optimal_fusion}
Let $\sigma^2_{c,u}$, $c=1,\ldots,C$, $u=1,\ldots,U$, be a set of given error variances for UE $u$ and cluster $c$.
Assume that the residual interference and noise terms are zero mean and uncorrelated among the clusters. 
Then, the weights that yield optimal fusion according to \fref{def:optfusion} are given by
\begin{align}
\nu_{c,u} = \frac{1}{\sigma_{c,u}^2}
\Bigg(\sum_{c'=1}^C \frac{1}{\sigma_{c',u}^2} \Bigg)^{\!\!-1},
\label{eq:opt_fusion}
\end{align}
for $c=1,\ldots,C$ and $u=1,\ldots,U$.
\end{lem}

\subsection{SINR Analysis of Optimal Fusion in the FD Architecture}
We are now interested in analyzing the post-fusion SINR for the FD architecture in the large-system limit.
The following theorem analyzes the decoupled noise variance $\sigma^2_c$ for each cluster $c=1,\ldots,C$; the proof is given in \fref{app:SE_FP_Decentralized}.

\begin{lem} \label{lem:SE_FP_Decentralized}
Assume MRC, ZF, L-MMSE, or LAMA equalization in each cluster $c=1,\ldots,C$. Consider the large-system limit and Rayleigh fading channels. Then, the input-output relation of each cluster is decoupled into parallel channels of the form~\fref{eq:decoupling} with decoupled noise variance~$\sigma_c^2$ given by a solution to the following fixed-point equation:
\begin{align*} 
w_c\sigma_c^2 = \No + \beta\Psi(\sigma_c^2).
\end{align*}
Here, $\Psi(\sigma_c^2)$ is the MSE function of the equalizer in cluster $c$.
\end{lem}

This result shows that the per-UE error variances $\sigma^2_{c,u}$ will become independent of the UE index~$u$ in the large-antenna limit and for Rayleigh-fading channels. Furthermore, the decoupled noise variances $\sigma_c^2$ depend on the fraction $w_c$ of BS antennas associated with cluster~$c$.

The following result establishes the post-fusion SINR in the FD architecture; a short proof is given in \fref{app:postfusionvariance}. 
\begin{thm} \label{thm:postfusionvariance}
Let the assumptions of \fref{lem:SE_FP_Decentralized} hold and $\sigma^2_{c}$, $c=1,\ldots,C$, be the per-cluster decoupled noise variances. Then, the decoupled noise variance $\sigma^2_\text{FD}$ of the fused estimate in \fref{eq:fusionrule} of the FD architecture is given by 
\begin{align}\label{eq:sigma_FD}
 \sigma^2_\text{FD} &=  
 \left(\sum_{c=1}^{\C} \frac{1}{\sigma_c^2}\right)^{\!\!-1}
\!=\No + \beta \sum_{c=1}^{\C} \nu_c \Psi(\sigma_c^2).
\end{align}
\end{thm}

We note that this result implies that the post-fusion SINR with optimal fusion according to \fref{def:optfusion}, denoted by $\SINR_\textnormal{FD}$, corresponds to the sum of the per-cluster SINR values. 

Finally, we have the following intuitive result which implies that for a given equalizer, the \FD{} architecture cannot outperform the \PD{} architecture; the proof is given in  \fref{app:LAMA_Arch2_Arch1}.
\begin{lem} Let $\No>0$ and assume the large-system limit and Rayleigh-fading channels. Then, the output 
SINR for the \FD{} and \PD{} architectures satisfy \mbox{$\SINR_\textnormal{FD} \leq \SINR_\textnormal{PD}$}. Equality holds for $\beta \to 0$, $C=1$, or if MRC-based equalization is used.\label{lem:LAMA_Arch2_Arch1}
\end{lem}

\subsection{Antenna Partitioning Strategies for Linear Equalizers}
We now analyze the post-fusion SINR performance of linear algorithms for the FD architecture, depending on the antenna allocation strategy, i.e., on the fraction of antennas $w_c$ used per cluster $c$. For the following analysis, we assume the large-system limit and Rayleigh fading channels. 

For MRC with the FD architecture, the post-fusion SINR is equivalent to that of the PD architecture (and that of centralized processing), as shown in \fref{lem:LAMA_Arch2_Arch1}. Hence, the antenna partitioning strategy does not affect the SINR performance. 

For ZF equalization in the FD architecture, we have the following result; the proof is given in \fref{app:ZF_same}.
\begin{lem}\label{lem:ZF_same}
Assume that the $B$ BS antennas are divided into~$C$ clusters so that $w_c \geq \beta$ holds for $c=1,\ldots,C$. 
Then, the post-fusion SINR for ZF equalization is given by
\begin{align}\label{eq:ZF_same}
\SINR_\textnormal{FD}^\textnormal{ZF} = \frac{\Es}{\No}(1 - C \beta).
\end{align}
\end{lem}
Interestingly, we observe that the post-fusion SINR $\SINR_\textnormal{FD}^\textnormal{ZF} $ does not depend on the antenna allocation strategy; this implies that the per-cluster antenna fraction $w_c$ can be chosen arbitrarily as long as $w_c \geq \beta$ for $c=1,\ldots,C$.\footnote{The condition $w_c \geq \beta$ implies that $\sum_{c=1}^C w_c = 1 \geq C\beta$ so the SINR expression in \fref{lem:ZF_same} is well-defined.}
Note, however, that equally-sized clusters are desirable in practice as they may minimize the interconnect or chip I/O bandwidth as well as the computational complexity per computing fabric. 

For L-MMSE equalization in the FD architecture, we have the following result; the proof is given in \fref{app:LMMSE_equalization}.
\begin{lem}\label{lem:LMMSE_equalization}
Assume that the $B$ BS antennas are divided into~$C$ clusters so that $\sum_{c=1}^Cw_c= 1$ holds with $w_c\geq 0$ for $c=1,\ldots,C$.
Then, the post-fusion SINR of the L-MMSE equalizer is given by
\begin{align}
\SINR_\textnormal{FD}^\textnormal{L-MMSE} =\,& 
\notag
\frac{1}{2}\sum_{c=1}^C
\sqrt{
\left(1 - \frac{\Es}{\No}(w_c-\beta)\right)^{\!\!2} + 4\frac{\Es}{\No}w_c}
\\
&- 
\frac{1}{2}\left(C - \frac{\Es}{\No}(1-C\beta)\right)\!
\label{eq:LMMSE_equalization}
\end{align}
\end{lem}

We see from \fref{lem:LMMSE_equalization} that the post-fusion SINR expression for L-MMSE equalization depends on the antenna allocation strategy, i.e., on the weights $w_c$, which is in contrast to ZF equalization (cf.~\fref{lem:ZF_same}). 
In addition, L-MMSE equalization does not require the restriction $w_c \geq \beta$ for ZF-equalization as the post-fusion SINR expression holds even  for underdetermined systems~\cite{TH1999}. 
Hence, it is natural to ask what the optimal  cluster allocation strategy is. 
The following result is rather disappointing; the proof is given in  \fref{app:LMMSE_upper}.

\begin{lem}\label{lem:LMMSE_upper}
The cluster allocation strategy that maximizes the post-fusion SINR for the L-MMSE equalizer $\SINR_\textnormal{FD}^\textnormal{L-MMSE}$ is $w_c=1$ for some $c$ and $w_{c'}=0$ for $c'\neq c$.
\end{lem}

Clearly, without any systematic requirements on the cluster ratios $w_c$, $c=1,\ldots,C$, besides $w_c\geq0$, maximizing $\SINR_\textnormal{FD}^\textnormal{L-MMSE}$ so that $\sum_{c=1}^C w_c =1$ corresponds to a centralized architecture, i.e., all antennas should be allocated to a single cluster.
Since the key idea of DBP was to mitigate interconnect and I/O bandwidth as well as computation bottlenecks, such an optimal allocation strategy is undesirable in practice. 
\revision{We next show that the most desirable (from a practical viewpoint) cluster allocation strategy, i.e., one for which all clusters have an equal number of antennas, yields the worst post-fusion SINR; the proof is given in \fref{app:LMMSE_uniform}.}
\begin{lem}\label{lem:LMMSE_uniform}
Assume that the $B$ BS antennas are divided into~$C$ clusters so that $\sum_{c=1}^Cw_c= 1$ with $w_c\geq 0$, $c=1,\ldots,C$.
Then, we have the following lower bound on the post-fusion SINR:
\begin{align}
\SINR_\textnormal{FD}^\textnormal{L-MMSE} \geq \, & 
\notag
\frac{1}{2}
\sqrt{
\left(1 - \frac{\Es}{\No}(1-C\beta)\right)^{\!\!2} + 4\frac{\Es}{\No}C}
\\
&- 
\frac{1}{2}\left(C - \frac{\Es}{\No}(1-C\beta)\right)\!.
\label{eq:LMMSE_lower}
\end{align}
Furthermore, the lower bound is achieved with equality if the antennas are distributed uniformly across all clusters, i.e., where $w_c = 1/C$ for  $c=1,\ldots,C$.
\end{lem}

We conclude by noting that even though uniform cluster sizes are the worst for L-MMSE equalization in the FD architecture, L-MMSE equalization outperforms ZF equalization for all possible partitioning schemes in terms of the post-fusion SINR. Hence, L-MMSE equalization is desirable in practice. 
%


\begin{figure*}[tp]
\centering
\subfigure[QPSK and target rate of $R=1.99$ {[bits/UE/channel\,use]}.]{
\includegraphics[width=0.7\columnwidth]{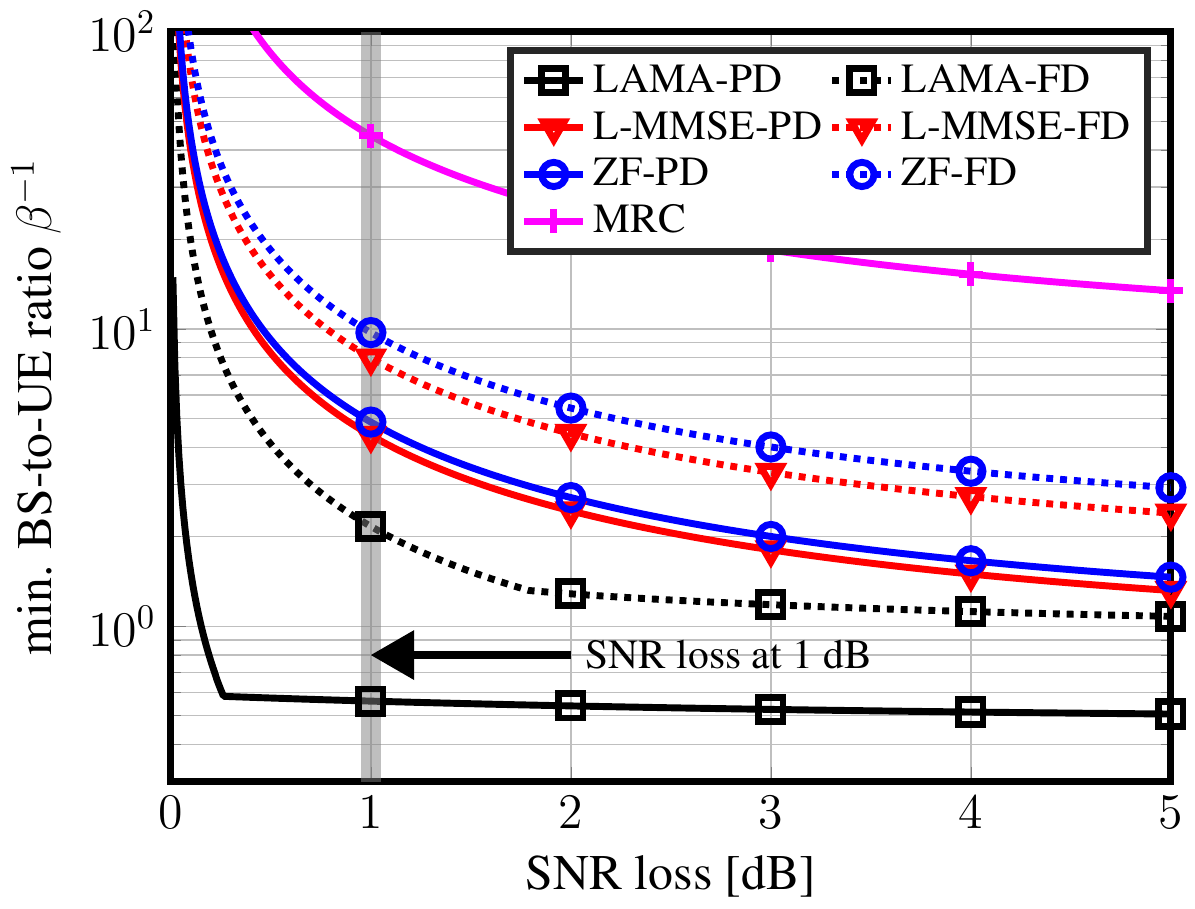}
\label{fig:fig_fixed_rate_QPSK}
}
\hspace{1.5cm}
\subfigure[16-QAM with target rate of $R=3$ {[bits/UE/channel\,use]}.]{
\includegraphics[width=0.7\columnwidth]{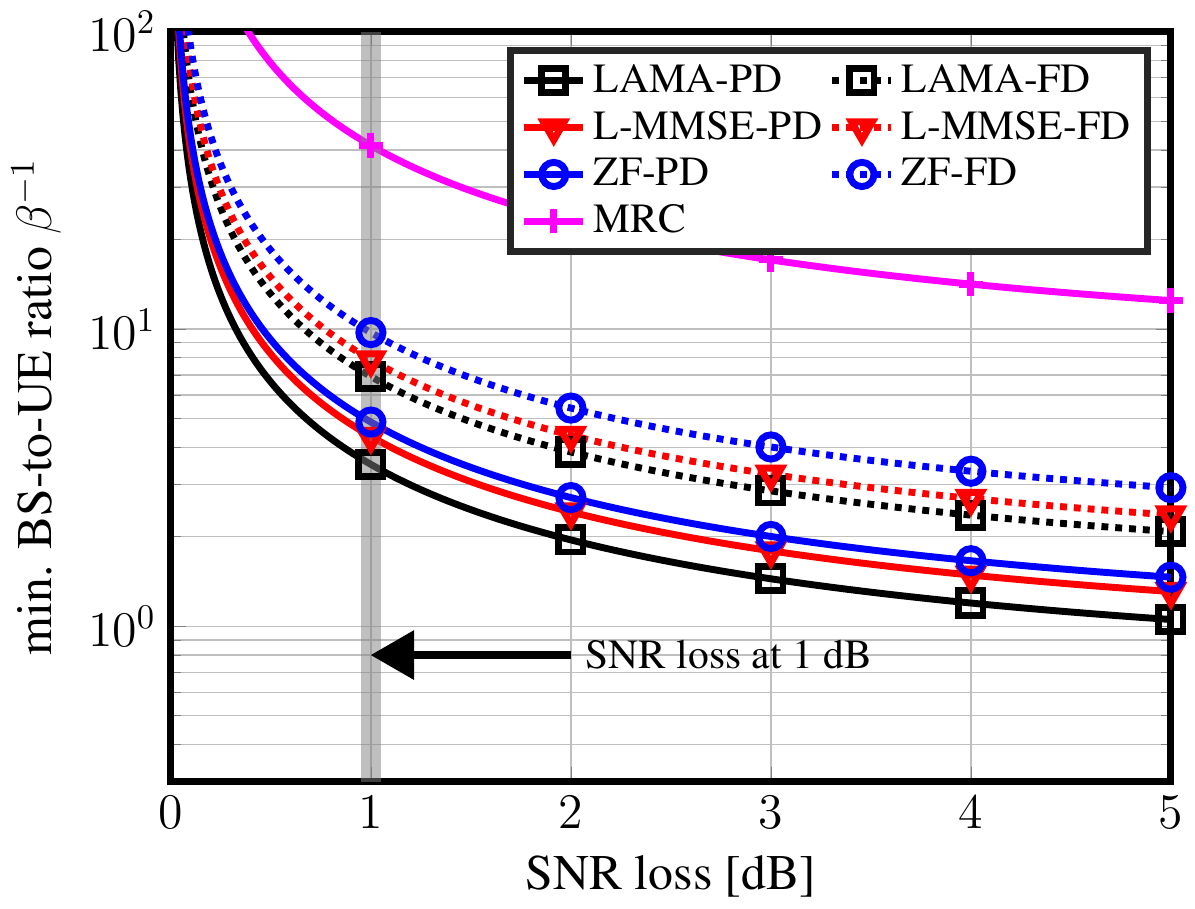}
\label{fig:fig_fixed_rate_16QAM}
}
\\
\hspace{0.1cm}
\subfigure[QPSK with fixed  1\,dB SNR loss.]{
\includegraphics[width=0.7\columnwidth]{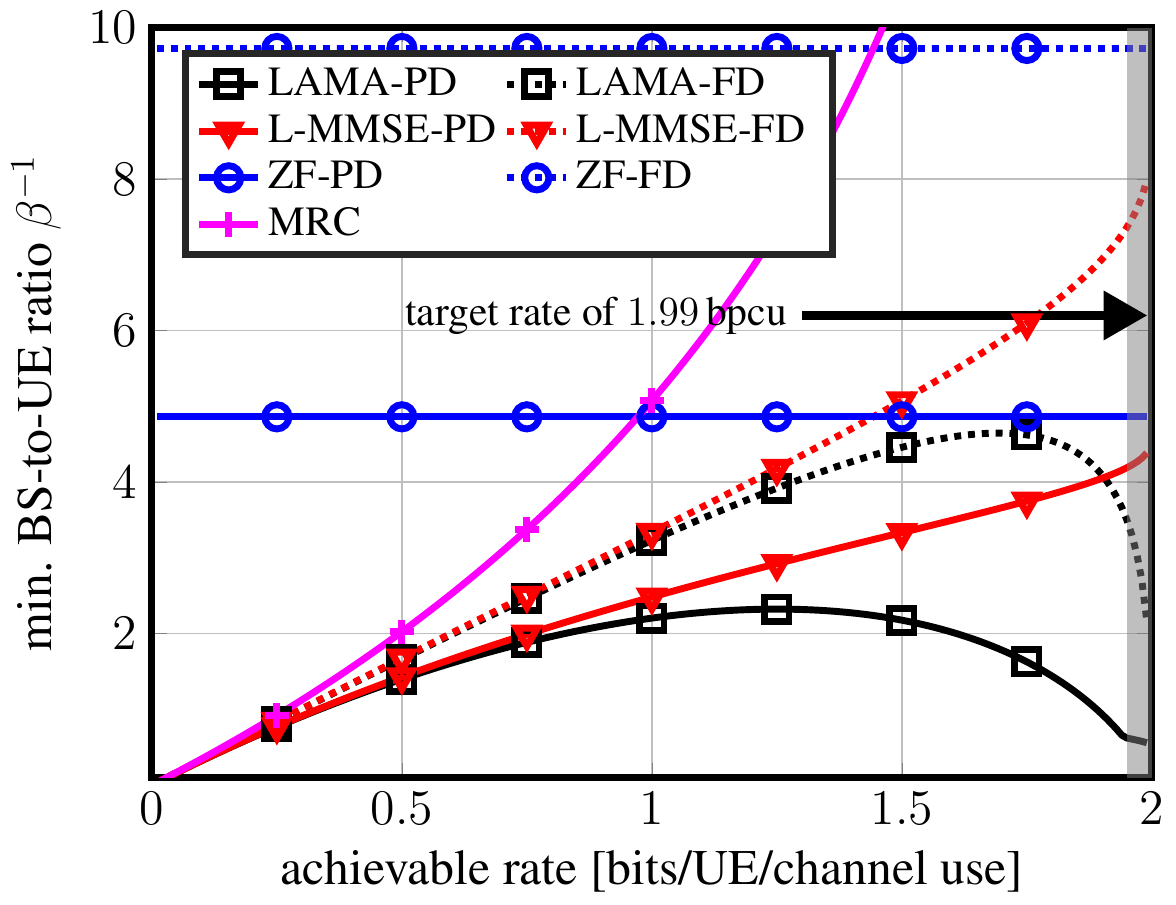}
\label{fig:fig_SNRloss_QPSK}
}
\hspace{1.5cm}
\hspace{0.05cm}
\subfigure[16-QAM with fixed  1\,dB SNR loss.]{
\includegraphics[width=0.7\columnwidth]{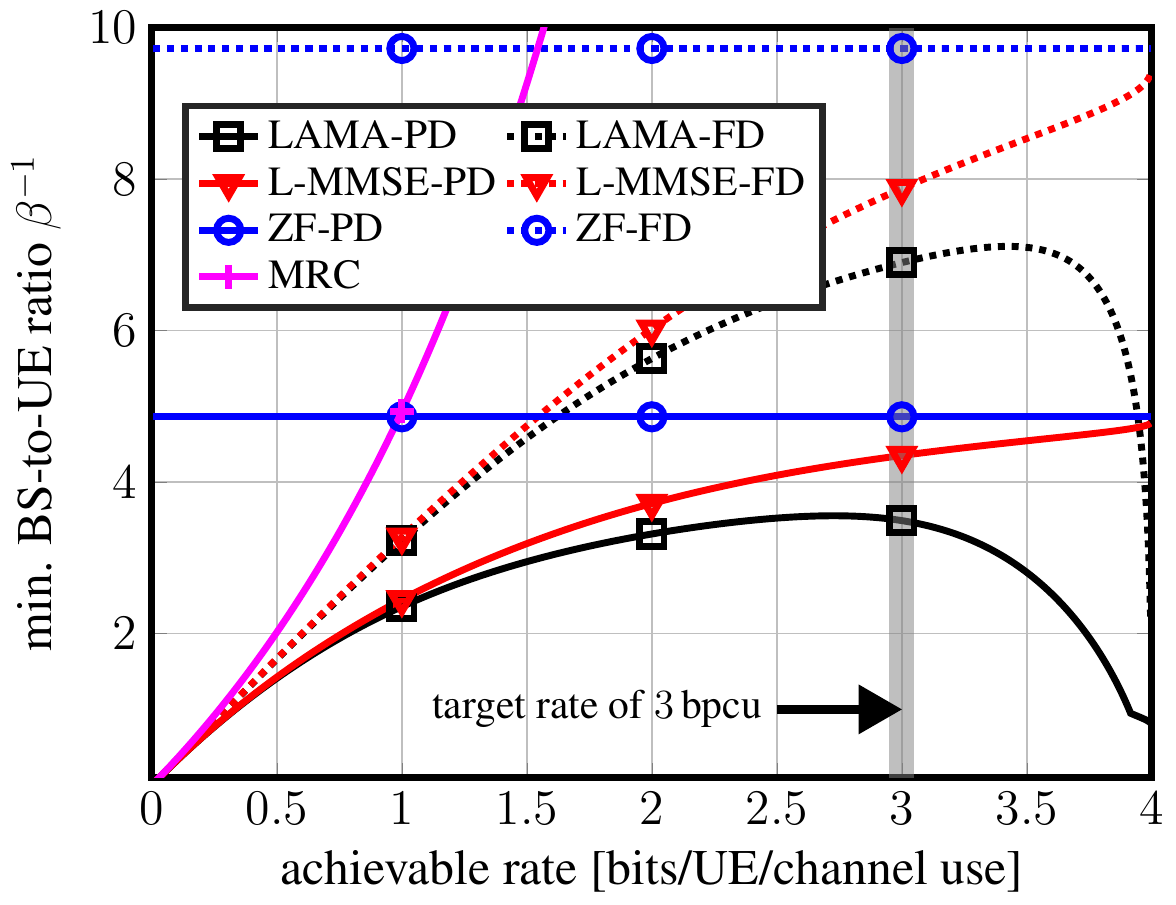}
\label{fig:fig_SNRloss_16QAM}
}
\centering
\caption{Achievable rate analysis of DBP with feedforward architectures. Minimum BS-to-UE antenna ratio $\beta^{-1}$ versus SNR loss for QPSK (a) and 16-QAM (b) for fixed target rates $R$. 
Minimum $\beta^{-1}$ versus achievable rate for QPSK (a) and 16-QAM (b) for a given SNR loss of $1$\,dB. 
LAMA outperforms MRC, ZF, and L-MMSE in the FD and PD architectures, especially at high target rates. At low target rates, all equalizers and  architectures perform equally well. 
}
\label{fig:numerical_results}
\end{figure*}

\section{Numerical Results} \label{sec:Results}
We now investigate the performance of decentralized equalization in the large-system limit and for Rayleigh-fading channels using the SINR expressions from Sections~\ref{sec:PDequalization} and~\ref{sec:FDequalization}. 
We show error-rate simulation results to validate our asymptotic results in finite-dimensional systems. We also provide results for an LTE-like massive MU-MIMO system to demonstrate the efficacy of our solutions in a more realistic scenario.

\subsection{Achievable Rate Analysis} \label{sec:Results_Rate}
We first investigate the achievable rates of our feedforward architectures with focus on the large-system limit and Rayleigh fading channels. We consider $C=2$ clusters with uniform antenna partitioning, i.e., $w_c=1/C$ for $c=1,\ldots,C$. 
We define the average receive signal-to-noise ratio (SNR) as \mbox{$\SNR=\beta\Es/\No$} and use an interference-free AWGN channel with variance~$\No$ as the baseline, which coincides to the large-antenna limit of massive MU-MIMO systems with $\beta\to0$. Concretely, we will use the following performance metric.

\begin{defi}[SNR loss] 
We define the SNR loss of an equalizer as the excess SNR required to achieve the same target rate~$R$ of an interference-free AWGN channel with variance $\No$. 
\end{defi}

In~\fref{fig:fig_fixed_rate_QPSK}, we use QPSK and a target rate of $R=1.99$ bits/UE/channel use; in~\fref{fig:fig_fixed_rate_16QAM} we use 16-QAM and a target rate of $R=3$ bits/UE/channel use. Both figures investigate the minimum required BS-to-UE ratio $\beta^{-1}$ for a given SNR loss, which characterizes how many more BS antennas are required by a given equalizer and feedforward architecture to be able to approach AWGN performance up to a given SNR gap. 
We observe that for a small SNR loss (i.e., when achieving similar performance as that of an interference-free AWGN channel), we require significantly more BS antennas than UEs, irrespective of the algorithm and architecture. 
For an SNR loss of $1$\,dB (shown by a thick vertical line in Figures~\ref{fig:fig_fixed_rate_QPSK} and~\ref{fig:fig_fixed_rate_16QAM}), we see that the PD architecture outperforms the FD architecture; this fact is more pronounced in the QPSK scenario as we are trying to achieve $99.5$\% of the maximum possible rate of~$2$ bits/UE/channel use for QPSK, whereas for 16-QAM, we are only trying to achieve $75$\% of the maximum rate. 
We also see that LAMA-PD significantly outperforms linear equalizers  in the \PD{} and \FD{} architectures; MRC requires significantly higher BS-to-UE antenna ratios. 
Interestingly, for QPSK, LAMA-\FD{} significantly outperforms linear equalization algorithms for the \PD{} architecture in \fref{fig:fig_fixed_rate_QPSK} but performs strictly worse for 16-QAM in \fref{fig:fig_fixed_rate_16QAM}; 
this is due to the fact that the system-ratio threshold for LAMA to achieve individually-optimal performance is higher for QPSK than for 16-QAM~\cite{JGMS2015conf}. 
In summary, LAMA-\FD{} achieves similar performance as linear equalizers with the \PD{} architecture while reducing interconnect and chip I/O bandwidths.

In~\fref{fig:fig_SNRloss_QPSK} and \fref{fig:fig_SNRloss_16QAM}, we fix the SNR loss to $1$\,dB and plot the minimum 
BS-to-UE ratio $\beta^{-1}$ and varying achievable rates for QPSK and 16-QAM, respectively. 
For both constellations, MRC performs equally well than all other methods in the low-rate regime; note that the low-rate regime translates to the low-SNR regime for which MRC is known to be optimal.
For higher rates, however, MRC requires significantly higher BS-to-UE antenna ratios compared to L-MMSE or LAMA-based equalization. 
The PD architecture significantly outperforms the FD architecture for all equalizers, which implies that for high-rates the PD architecture is the preferred choice. 
Interestingly, the minimum BS-to-UE antenna ratio remains constant for ZF; this implies that as long as one operates below a certain antenna ratio $\beta^*$, ZF is able to support all transmission rates; see \cite{ghods2017optimally} for additional details on this behavior. 
Finally, we see that the minimum BS-to-UE ratio $\beta^{-1}$ decreases for LAMA-\FD{} and LAMA-\PD{} at high rates; this behavior is due to the fact that LAMA in overloaded systems is particularly robust at low and high values of SNR (see \cite{JGMS2015conf} for a detailed discussion).

\begin{figure}[tp]
\centering
\includegraphics[width=0.72\columnwidth]{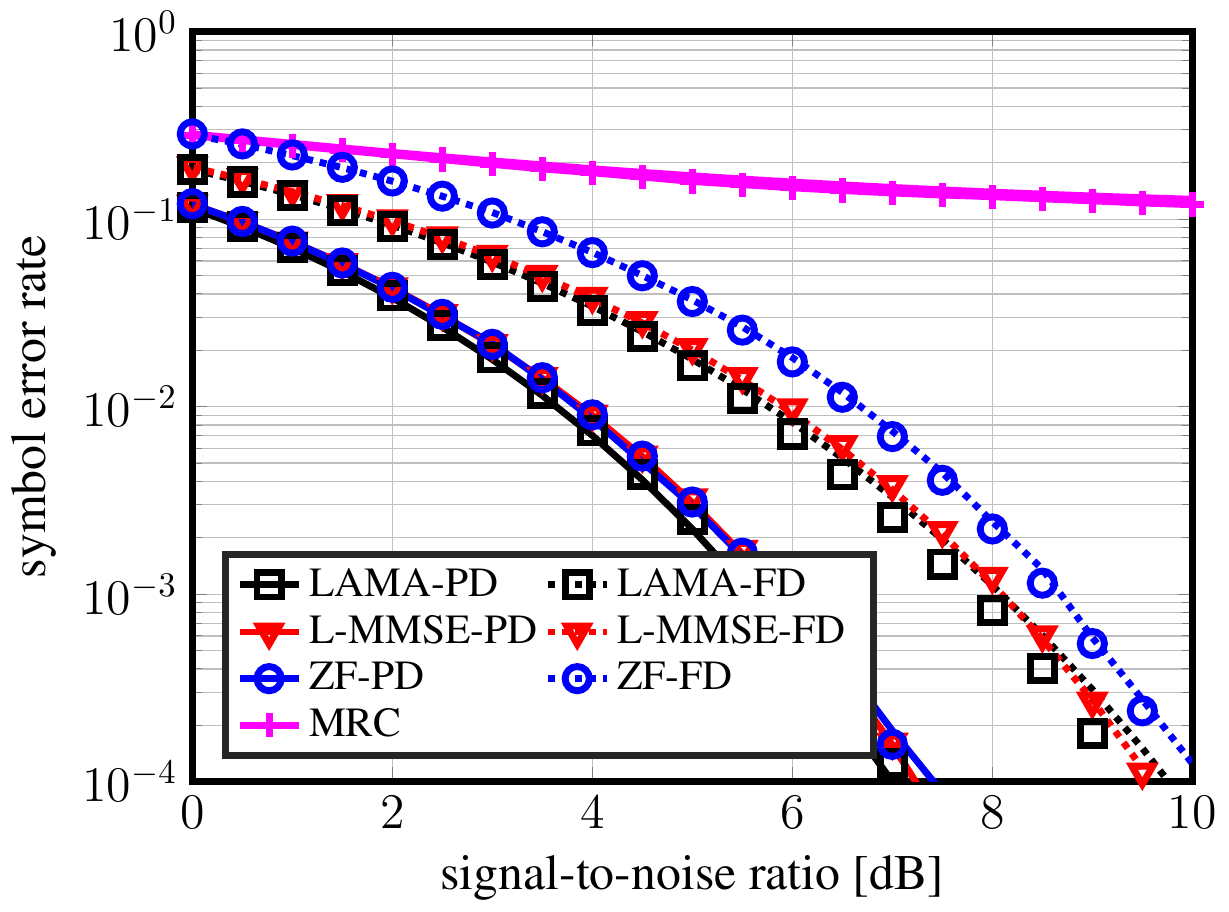}
\vspace{-0.1cm}
\caption{Symbol error-rate (SER) comparison of equalization in the large-antenna limit (indicated with lines) vs. the simulated performance (indicated by the markers) in a $B=256$ BS antenna $U=16$ UE massive MU-MIMO system with Rayleigh-fading channels. Evaluating the SER using our analytical SINR expressions for the large-antenna limit closely matches numerical simulations in finite-dimensional systems for all considered equalizers and  architectures.}
\label{fig:asymptoticvsfinite}
\end{figure}

\subsection{Asymptotic vs.\ Finite-Dimensional Systems} \label{sec:Results_SERSE}
In \fref{fig:asymptoticvsfinite}, we compare our analytical SINR expressions in the large-antenna limit to those in a finite-dimensional massive MU-MIMO scenario. 
Specifically, we use the SINR from \fref{eq:decoupling} in a decoupled AWGN channel to analytically compute the symbol error-rate (SER) as well as the simulated SER in an uncoded $B=256$ BS antenna, $U=16$ UE massive MU-MIMO system with Rayleigh fading channels. 
We consider $C=8$ clusters with uniform antenna partitioning, i.e., $w_c = 1/8$ for all $C=8$ clusters.
First, we observe that the simulated results (indicated with markers) closely match our analytical expressions (indicated with lines). 
Second, we see that the PD architecture significantly outperforms the FD architecture for $C=8$ clusters and LAMA outperforms ZF and L-MMSE for both architectures. 
Third, we see that MRC yields poor SER performance, which is due to the fact that MRC requires extremely high BS-to-UE antenna ratios to support $4$\,bits/UE/channel use for 16-QAM; see also~\fref{fig:fig_SNRloss_16QAM}.
%

\begin{figure}[tp]
\centering
\includegraphics[width=0.7\columnwidth]{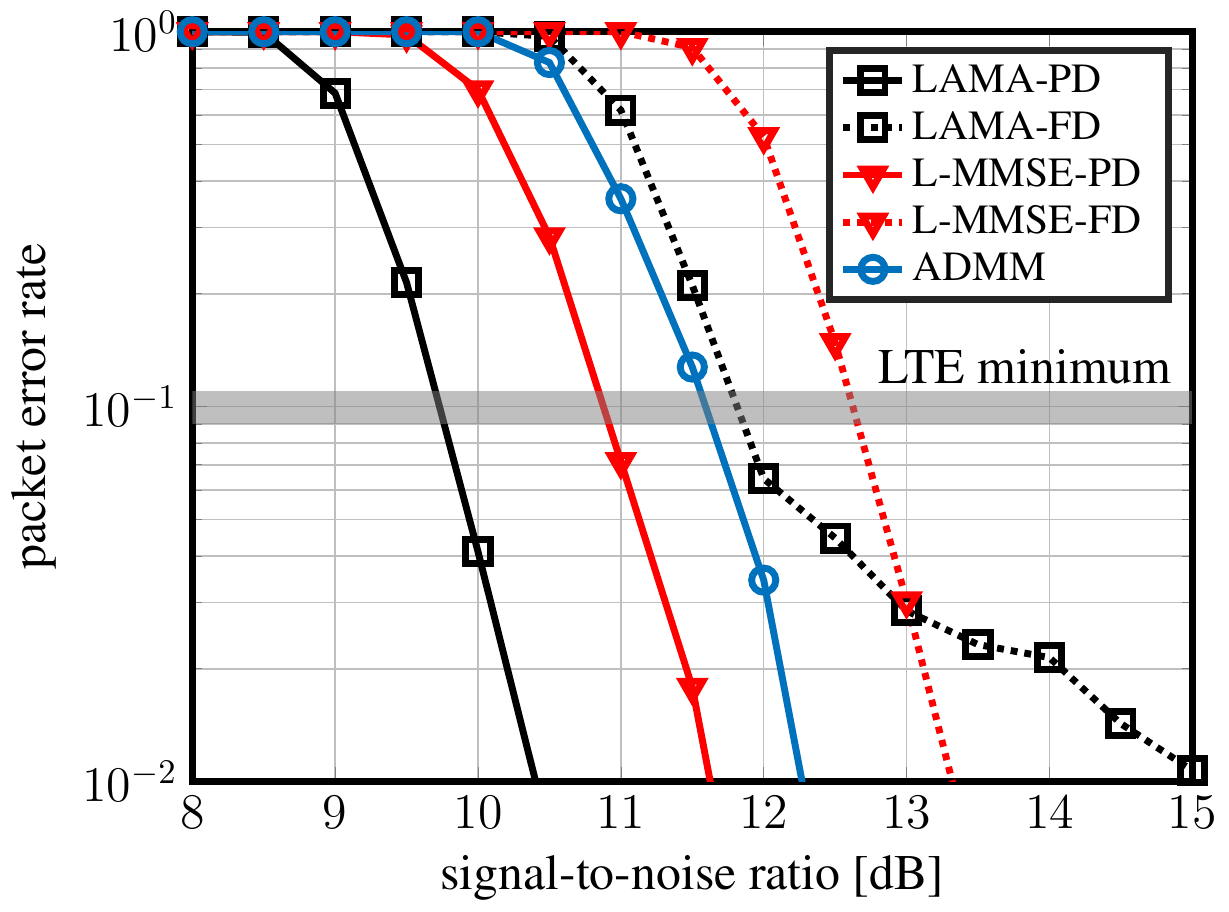}
\label{fig:64x16}
\vspace{-0.1cm}
\caption{Packet error-rate (PER) of an LTE-like massive MU-MIMO-OFDM system with $B=64$, $U=16$, \revision{and 56\,k bits per OFDM packet. }
LAMA for the PD and FD architectures clearly outperforms L-MMSE while meeting the $10$\% LTE minimum PER requirement. 
LAMA-PD or L-MMSE clearly outperform the consensus-based ADMM method from \cite{li2017decentralized} (which suffers from transfer latency), whereas LAMA-FD closely approaches the performance at minimal lower latency overheads.
}
\label{fig:LTE_error_rate_analysis}
\end{figure}

\subsection{Coded Error-rate Performance in Realistic Systems} \label{sec:Results_codedBER}

\revision{Since our theoretical analysis relies on the large system limit and Rayleigh fading channels, we now investigate the efficacy of our work in a more realistic scenario.} 
Specifically, \fref{fig:LTE_error_rate_analysis} shows the coded packet error-rate (PER) in a realistic LTE-like massive MU-MIMO system with $B=64$ BS antennas and $U=16$ UEs. 
We consider $C=2$ clusters with uniform antenna partitioning.
We use OFDM with $2048$ subcarriers ($1200$ used for data transmission) with 16-QAM, $14$ OFDM symbols per packet, and use a weak rate-$5/6$ convolutional code with soft-input Viterbi decoding. 
We consider a WINNER II channel model in an outdoor-to-indoor scenario. 
For LAMA-PD and FD, we use $10$ iterations and perform message damping to mitigate performance loss for finite-dimensional systems with correlated MIMO channel matrices~\cite{VSRKZ2015}.
\revision{
We observe that LAMA-FD exhibits an error floor near a PER of $10^{-2}$ due to the small effective system dimension compared to that given by LAMA-PD. Nevertheless,  LAMA-FD still outperforms L-MMSE-FD for the LTE minimum PER specification of 10\%.}

We also compare LAMA and L-MMSE to the consensus-based ADMM method for DBP proposed in \cite{li2017decentralized}, where we use $10$ iterations. 
First, we see that LAMA-\PD{} outperforms all other equalization algorithms by a significant margin, when considering the LTE minimum PER specification of $10$\%.
Second, we observe that the consensus-sharing ADMM method performs slightly better than that of LAMA-\FD{}. The ADMM-based method, however, requires iterative consensus exchange among the clusters which results in low throughput; see our GPU implementation results in \fref{sec:GPUperformance}.
%


\section{Multi-GPU System Implementation}
\label{sec:gpuimplementation}

\begin{figure*}[tp]
\centering
\subfigure[t][$C=1$]{\includegraphics[height=0.25\linewidth]{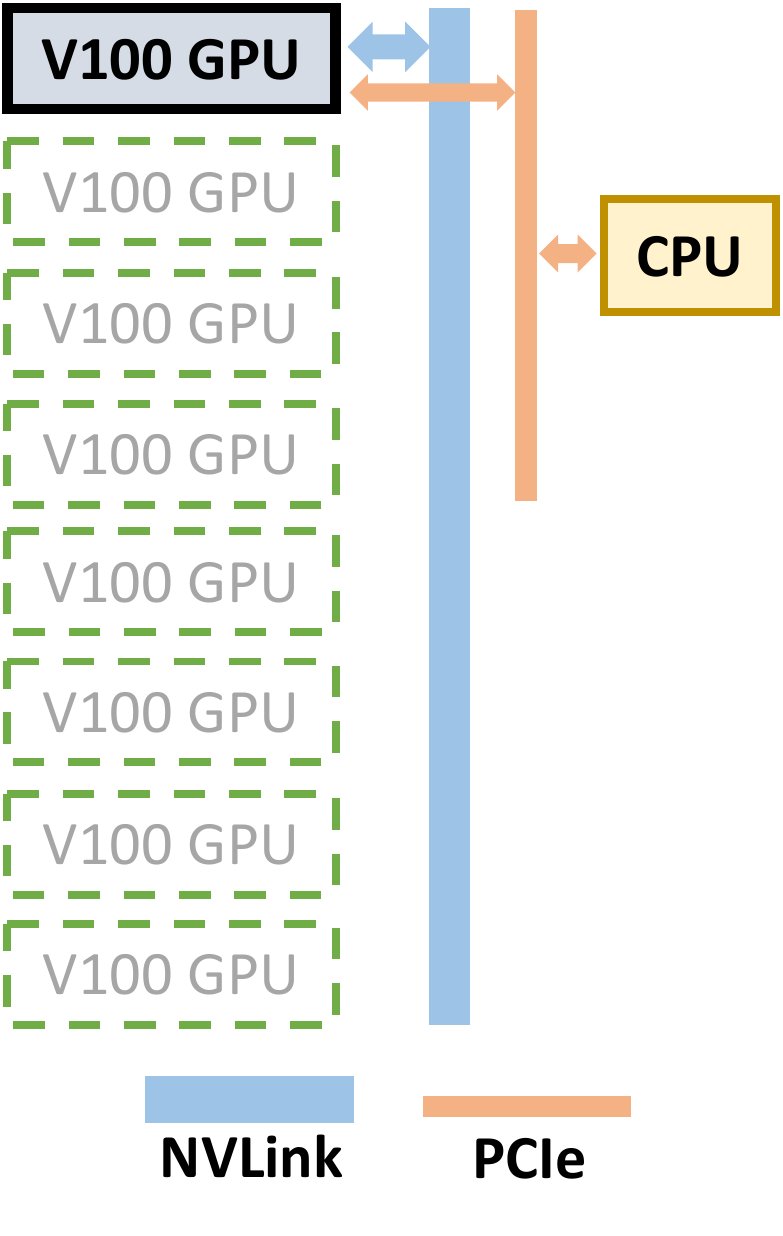}\label{fig:arch_c1}}
\hspace{0.9cm}
\subfigure[t][$C=2$]{\includegraphics[height=0.25\linewidth]{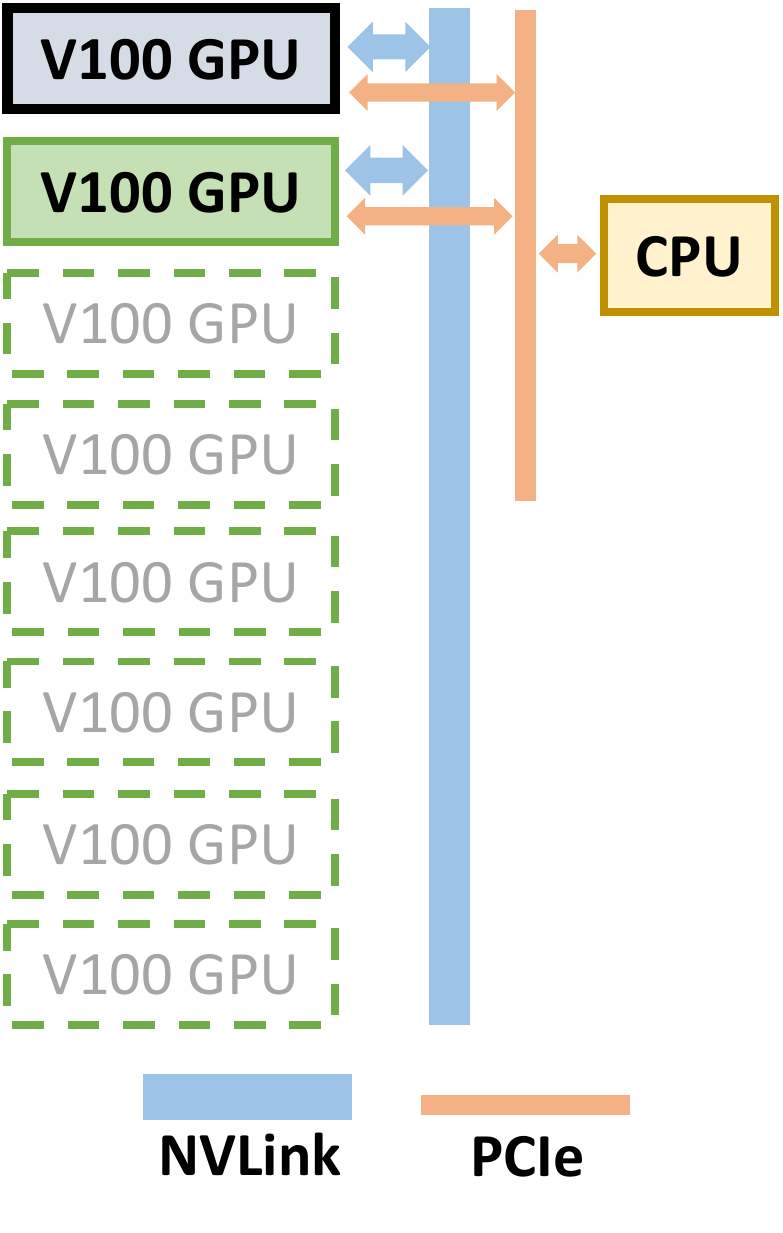}\label{fig:arch_c2}}
\hspace{0.9cm}
\subfigure[t][$C=4$]{\includegraphics[height=0.25\linewidth]{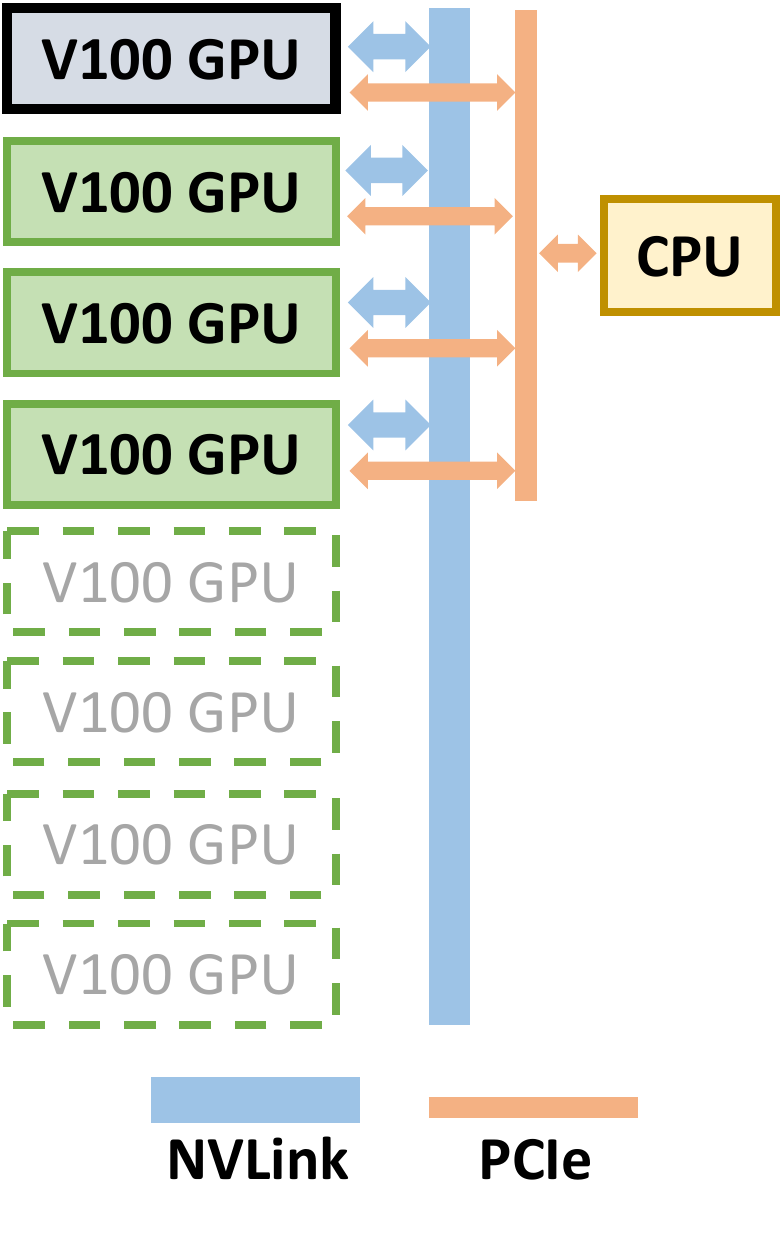}\label{fig:arch_c4}}
\hspace{0.9cm}
\subfigure[t][$C=8$]{\includegraphics[height=0.25\linewidth]{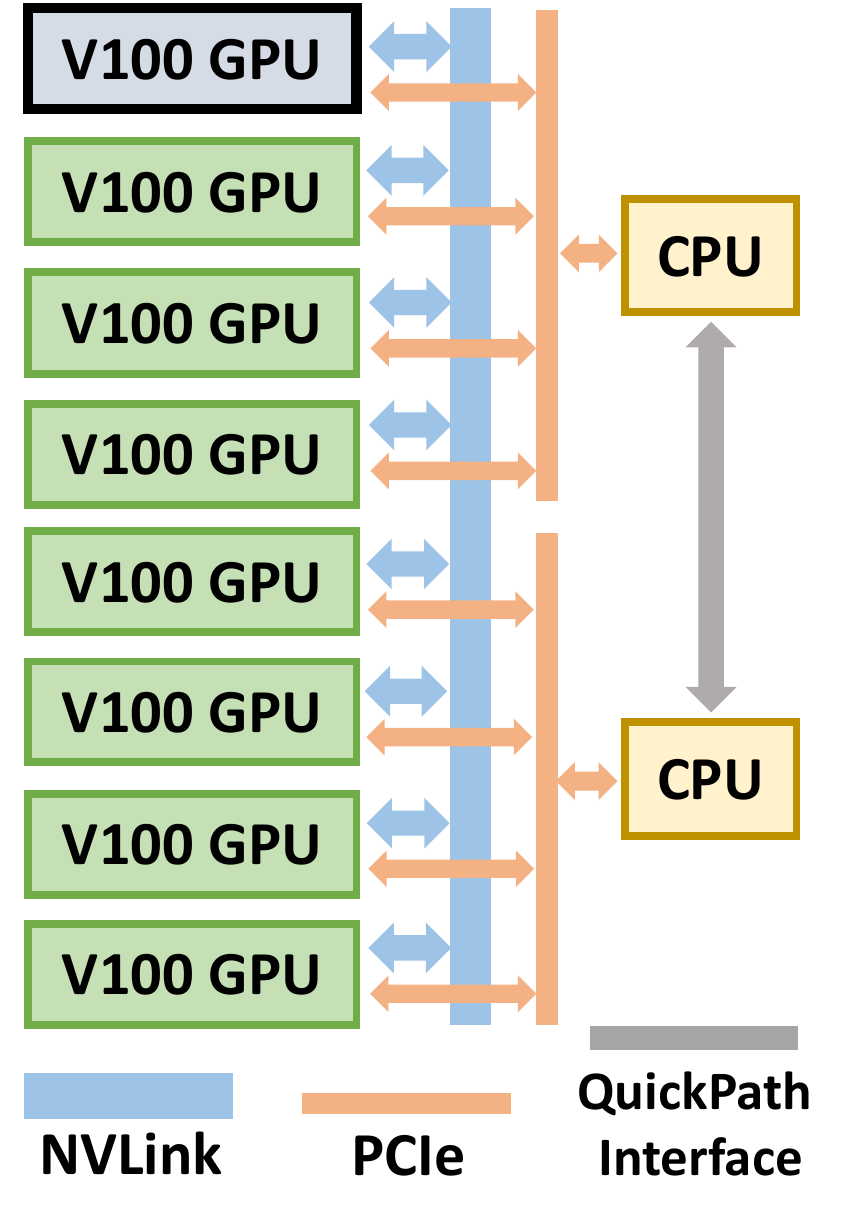}\label{fig:arch_c8}}
\caption{System architecture overview of the multi-GPU experimental platform. The system consists of eight Tesla Volta GPUs with high-bandwidth NvLink GPU-to-GPU interconnect. The master GPU (highlighted) gathers local equalization results and performs all centralized computations. For (a) $C=1$, (b) $C=2$, and (c) $C=4$ clusters, we use a single CPU that controls the GPUs via PCIe; (d) for $C=8$ clusters, we use two CPUs for control purposes. }
\label{fig:sys_arch}
\end{figure*}

We now present implementation results of our algorithms and architectures on a multi-GPU system to validate the scalability, throughput, and latency in a realistic scenario. We furthermore provide a comparison to existing consensus-based DBP algorithms and centralized equalizers. 
In order to fully utilize the available GPU resources, we consider an OFDM-based system as in \fref{sec:Results_codedBER}, which enables us to not only parallelize across antennas but also across subcarriers. 

\begin{remark}
As in  \cite{li2017decentralized}, the provided multi-GPU implementations serve as a proof-of-concept
to assess the efficacy and scalability of our solutions when implemented on real hardware.
In practice, we expect that our solutions will be implemented on clusters consisting of field-programmable gate arrays (FPGAs) or application-specific integrated circuits (ASICs), which offer higher computing efficiency and lower transfer latency. 
\end{remark}

\subsection{Experimental Platform}
In Fig. \ref{fig:sys_arch}, we show the used Nvidia DGX-1 multi-GPU system~\cite{dgx1} consisting of eight Tesla V100 Volta GPUs with  
$300$\,GB/s bi-directional NvLink interconnect and two $20$-core Intel Xeon E5-2698 v4 CPUs. Each GPU provides 5120 CUDA cores and  
16\,GB high bandwidth memory (HBM). 
In what follows, we use $C$ GPUs when processing $C$ antenna clusters; other partitioning schemes are possible.
Our equalization algorithms and architectures are implemented with the CUDA~9.2 toolkit~\cite{cuda} and the Nvidia collective communications library~(NCCL) v2.2~\cite{nccl} software stack. 
The NCCL uses the message passing interface (MPI) library, which improves the efficiency  of inter-GPU collective communication over NvLink~\cite{nvlink} by leveraging the CUDA-aware MPI technology~\cite{cudaaware} for direct GPU-to-GPU memory copy.

\begin{table*}[tp]

\caption{Latency (L) and throughput (TP) performance of decentralized feedforward equalizers ($U=16$, $B_c=32$, L in ms, TP in Gb/s)}
\renewcommand{\arraystretch}{1.05}
\centering
\begin{tabular}{@{}lcccccc@{}}
\toprule 
&\multicolumn{3}{c}{Partially Decentralized \revision{(PD)} Feedforward Equalizers} & \multicolumn{3}{c}{Fully Decentralized \revision{(FD)} Feedforward Equalizers }\tabularnewline
\midrule 
$B$ & 64 & 128 & 256 & 64 & 128 & 256\tabularnewline
$C$ & 2 & 4 & 8 & 2 & 4 & 8\tabularnewline
\midrule
Performance & L / TP  & L  / TP  & L  / TP  &   L  / TP  & L  / TP  & L  / TP\tabularnewline
\midrule 
LAMA, $T_\text{max}=1$ & 0.651 / 1.76 & 0.677 / 1.69 & 0.785 / 1.46 &  0.591 / 1.94 & 0.599 / 1.91 & 0.672 / 1.71\tabularnewline
LAMA, $T_\text{max}=2$ & 0.777 / 1.48 & 0.803 / 1.43 & 0.912 / 1.26 &  0.718 / 1.60 & 0.727 / 1.58 & 0.799 / 1.44\tabularnewline
LAMA, $T_\text{max}=3$ & 0.903 / 1.27 & 0.929 / 1.23 & 1.039 / 1.10 &  0.846 / 1.36 & 0.854 / 1.34 & 0.925 / 1.24\tabularnewline
LAMA, $T_\text{max}=4$ & 1.030 / 1.11 & 1.056 / 1.09 & 1.165 / 0.98 &   0.973 / 1.18 & 0.980 / 1.17 & 1.052 / 1.09\tabularnewline
LAMA, $T_\text{max}=5$ & 1.156 / 0.99 & 1.183 / 0.97 & 1.291 / 0.89 &   1.099 / 1.04 & 1.106 / 1.03 & 1.180 / 0.97\tabularnewline 
L-MMSE & 0.719 / 1.60 & 0.745 / 1.54 & 0.856 / 1.34  & 0.666 / 1.72 & 0.676 / 1.70 & 0.751 / 1.53 \tabularnewline
ZF & 0.690 / 1.66 & 0.717 / 1.61 & 0.827 / 1.39 &  0.635 / 1.81 & 0.643 / 1.79 & 0.715 / 1.61\tabularnewline
MRC & 0.411 / 2.79 & 0.421 / 2.72 & 0.499 / 2.30 &  0.411 / 2.79 & 0.421 / 2.72 & 0.499 / 2.30 \tabularnewline
\bottomrule 
\end{tabular}
\label{tbl:gpuresult}
\end{table*}

\begin{table}[tp]
\caption{Performance of decentralized consensus-sharing equalizers developed in \cite{li2017decentralized}  ($U=16$,  $B_c=32$,  L in ms, TP in Gb/s)}
\renewcommand{\arraystretch}{1.05}
\centering
\begin{tabular}{@{}lccc@{}}
\toprule 
$B$ & 64 & 128 & 256 \tabularnewline
$C$ & 2 & 4 & 8 \tabularnewline  
\midrule
Performance & L / TP  & L  / TP  & L  / TP \tabularnewline
\midrule
ADMM, $T_\text{max}=2$ & 0.783 / 1.47 & 0.789 / 1.45 & 0.858 / 1.34 \tabularnewline  
ADMM, $T_\text{max}=3$  & 0.982 / 1.17 & 0.995 / 1.15 & 1.075 / 1.07 \tabularnewline
CG, $T_\text{max}=2$   & 0.808 / 1.42 & 0.815 / 1.41 & 0.880 / 1.30 \tabularnewline
CG, $T_\text{max}=3$  & 0.997 / 1.15 & 1.010 / 1.14 & 1.098 / 1.04 \tabularnewline
\bottomrule 
\end{tabular}
\label{tbl:gpuconsensus}
\end{table} 

\begin{table}[tp]
\centering
\caption{Performance of centralized equalizers ($U=16$, $C=1$, $B_c=B$,  L in ms, TP in Gb/s)}
\renewcommand{\arraystretch}{1.05}
\centering
\begin{tabular}{@{}lccc@{}}
\toprule 
$B$ & 64 & 128 & 256 \tabularnewline 
\midrule
Performance & L / TP  & L  / TP  & L  / TP \tabularnewline
\midrule
LAMA, $T_\text{max}=2$ & 0.766 / 1.50 & 1.182 / 0.97 & 2.004 / 0.57\tabularnewline  
L-MMSE~\cite{Li_DCAS15}  & 0.710 / 1.62 & 1.125 / 1.02 & 1.947 / 0.59\tabularnewline
ZF~\cite{Li_DCAS15}  & 0.682 / 1.68 & 1.095 / 1.05 & 1.917 / 0.60\tabularnewline
MRC  & 0.457 / 2.51 & 0.859 / 1.34 & 1.650 / 0.70\tabularnewline
\bottomrule 
\end{tabular}
\label{tbl:gpucentral}
\end{table}

\subsection{Implementation Details}
All decentralized feedforward equalizers proposed in Sections~\ref{sec:PDequalization} and~\ref{sec:FDequalization} are processed in the following three steps:
(i) calculate local intermediate results at each antenna cluster,
(ii) fuse local results at the centralized processor, and
(iii) perform necessary centralized computations to obtain the final equalization outputs. 
We use the available CPUs to schedule the workload and initialize $C$ MPI processes, 
each process supervising an individual GPU for design decentralization. 
The proposed decentralized algorithms were implemented on the multi-GPU system using the following procedure: 
%
(i) We accelerate local computations at the $C$ antenna clusters, each using a dedicated GPU with multi-threaded CUDA kernel functions.
%
(ii) We utilize collective inter-process communication among the $C$ GPUs via NCCL over NvLink to realize low-latency gathering of the local results at the master GPU.
%
(iii) We complete the centralized computations at the master GPU for high efficiency. 
%
%
We note that equalization with the PD architecture generally requires fewer local computations, but more data for fusion and computations at the centralized master GPU, compared to equalization with the FD architecture. We now describe the \PD{} and \FD{} architectures in detail. To keep the discussion of the FD architecture short, we only discuss the procedure for LAMA, as the same methodology is used for the linear equalization algorithms. \revision{In order to minimize the complexity of computing the post-equalization variance and SINR~\cite{studer2011asic}, our implementations resort to the low-complexity approximation put forward in~\cite{Li_DCAS15}.}

\subsubsection{PD Architecture}
%
The local computations at each GPU include the partial MRC output $\bmymrc_c=\bH_c\bmy_c$ and the partial Gram matrix $\bG_c = \bH^\Herm_c\bH_c$.
To maintain high GPU occupancy, we aggregate these workloads across a batch of subcarriers and process them in parallel.
To process the batched matrix-matrix or matrix-vector multiplications required for partial MRC and Gram computations efficiently, we take advantage of the \emph{cuBLAS} library, a collection of CUDA-accelerated basic linear algebra subprograms (BLAS), and specifically, the \texttt{cublasCgemmBatched} library function for fast matrix multiplications with complex-valued floating-point entries. 
In a single invocation of the batched function call, we calculate $\bG_c$ for $N_\textnormal{sc}$ active subcarriers ($\texttt{batchsize}=N_\textnormal{sc}$) associated with a certain OFDM symbol, and reuse them across $N_\textnormal{sym}$ OFDM symbols within the channel coherence time to reduce complexity. 
In contrast to the Gram matrix calculation, we compute $\bmymrc_c$ for a total of $N_\textnormal{sc}\times N_\textnormal{sym}$ subcarriers ($\texttt{batchsize}=N_\textnormal{sc}\times N_\textnormal{sym}$) in a function call. 
This is necessary because $\bmymrc_c$ also depends on the receive vector~$\bmy_c$, which varies for each OFDM symbol. 
We finally fuse the local $\bG_c$ and $\bmymrc_c$ that results in a total message size \revision{$m_\text{PD}=U^2 \times N_\textnormal{sc}\times C+U\times N_\textnormal{sc}\times N_\textnormal{sym}\times C$} at the master GPU \revision{with the \texttt{ncclReduce} NCCL function.} 

For LAMA-\PD{} shown in \fref{alg:lamanew}, each iteration mainly involves matrix-vector multiplication and vector operations, which include vector addition, subtraction, and dot-products.
Although matrix-vector multiplications can be efficiently computed by the batched {cuBLAS} function with $\texttt{batchsize}=N_\textnormal{sc}\times N_\textnormal{sym}$, as done for preprocessing, we also designed customized multi-threaded kernel functions for the vector operations to fully exploit the on-chip memories of GPU. 
Specifically, for certain kernels, we combine multiple vector operations which can share intermediate results using fast local registers. 
Since the intermediate vectors within each LAMA iteration, such as $\bms$, $\bmv$, and $\bmz$, scale with the UE number $U$, multi-threaded computations for vector additions, subtractions, and scaling are straightforward; we launch a kernel with $N_\textnormal{sc}\times N_\textnormal{sym} \times U$ threads to process each of $U$ entries in a vector for a total of $N_\textnormal{sc}\times N_\textnormal{sym}$ subcarriers in parallel. 
For the vector dot-product operations, as an example, when we update~$\phi$ in \fref{alg:lamanew}, we resort to the \emph{warp shuffle} technique, where a thread can directly retrieve register values from another thread efficiently, to obtain a sum reduction of~$U$ vector entries across~$U$ threads in the same \emph{warp}. 
If $U>\texttt{warpsize}=32$, then we can also use the on-chip shared memory for the necessary inter-thread communication. After the last LAMA iteration $T_\text{max}$, we finally calculate the global LAMA-PD output at the master GPU.

\subsubsection{FD Architecture}
Preprocessing and equalization is computed in a decentralized manner at each of the $C$ GPUs. 
We reuse the \emph{cuBLAS} library and customized kernel functions for those local computations in LAMA-\FD{}, and fuse the local result $\bmz_c$ at the master GPU using the NCCL \revision{\texttt{ncclReduce} function} with a smaller total message size \revision{$m_\text{FD}=U\times N_\textnormal{sc}\times N_\textnormal{sym}\times C$} than that of LAMA-PD. 
We implement optimal fusion at the master GPU.

\subsection{Implementation Results}
\label{sec:GPUperformance}

We now present measured latency and throughput of our decentralized feedforward equalizers. In \fref{tbl:gpuresult}, we list results \revision{of different equalizers with PD and FD architectures. These results are} obtained via CPU wall clock timing that is synchronized with the start and end of all GPU computations.
In what follows, we consider 16-QAM, and we fix the number of UEs to $U=16$ and BS antennas per cluster to $B_c=32$
We simulate an LTE-like system as in \fref{sec:Results_codedBER} with $N_\textnormal{sym}=14$ OFDM symbols per packet and $N_\textnormal{sc}=1200$ active subcarriers (out of 2048 subcarriers), which corresponds to a 20\,MHz LTE subframe. 
To compare the scalability of our proposed feedforward equalization architectures, we scale the total number of BS antennas $B=CB_c$ by increasing the number of clusters from $C=2$, $C=4$, to $C=8$. 

\subsubsection{Throughput and Latency}
We see from~\fref{tbl:gpuresult}  that all of our proposed  decentralized feedforward equalizers achieve throughput in the Gb/s regime with latencies of $1$\,ms or less.
Specifically, the non-linear LAMA-PD and LAMA-FD equalizers are able to reach higher throughput for a small number of iterations ($T_\text{max}=1$ or $T_\text{max}=2$). 
We note that for more LAMA iterations, our decentralized linear equalizers
are able to outperform LAMA in terms of the throughput.
This is due to the fact that linear equalizers can reuse both local Gram matrix multiplication and matrix inversion results across $N_\textnormal{sym}$ OFDM symbols, which effectively reduces complexity.
Among all those feedforward equalizers, MRC (which is equivalent for MRC-\PD{} and MRC-\FD{}) has the highest throughput and lowest latency, but entails a significant error-rate performance loss; see the simulation results in \fref{sec:Results}.
\revision{Furthermore, we see that equalization with the FD architecture generally achieves higher throughput than that of the PD architecture, mainly caused by smaller message sizes and lower data-transfer latency during the data fusion process. This advantage, however, comes at the cost of reduced error-rate performance for the FD architecture. 
Put simply, there exists a trade-off between throughput and error-rate performance for the FD and PD architectures.
}

\revision{Interestingly, the latency and throughput of our decentralized feedforward equalization implementations degrades only slightly when we increase the number of BS antennas (with a larger number of clusters $C$) as our designs benefit from direct GPU-to-GPU communication with efficient NCCL over NvLink.} This observation demonstrates that our proposed decentralized feedforward equalizers have excellent scalability to support hundreds to even thousands of BS antennas while maintaining high throughput.

We now compare our proposed decentralized feedforward architectures to the decentralized consensus-based methods proposed in~\cite{li2017decentralized}.
In \fref{tbl:gpuconsensus}, we show reference latency and throughput results measured on the same multi-GPU platform.
As discussed in~\cite{li2017decentralized}, consensus-sharing methods, such as ADMM-based and decentralized CG-based equalizers, rely on iterative update of local variables and global consensus, which requires the data gathering for consensus calculation and consensus broadcasting \emph{in each iteration}. 
In contrast, the decentralized feedforward equalizers proposed in this paper significantly reduce the data transfer latency with only \emph{one-shot} (feedforward) message passing, which leads to (often significantly) higher throughput. 
More specifically, both LAMA-PD and LAMA-FD achieve higher throughput than D-ADMM or D-CG for the same number of iterations; the same trend continues to hold for a larger number of iterations. 

In~\fref{tbl:gpucentral}, we show the latency and throughput performance of several \emph{centralized} equalizers. We see that the throughput  of centralized equalizers decreases quickly when increasing the number of BS antennas; this demonstrates that centralized solutions exhibit poor scalability to large BS antenna arrays. 
Also note that centralized solutions suffer from high interconnect and chip I/O data rates, which is not visible in this comparison. 

\subsubsection{\revision{Data Transfer}}
\revision{The key advantages of the proposed decentralized feedforward architectures are as follows: (i) They reduce the raw front-end baseband transfer rate by a factor of $C$ at each antenna cluster compared to that of centralized architectures and (ii) they minimize the back-end message passing data among clusters. For example, given $N_\text{sc}=1200$, $N_\text{sym}=14$, $U=16$ and $C=4$, the total message passing data across $C$ GPUs with the PD architecture is $m_\text{PD} = 17.58$\,MB while the FD architecture requires only  $m_\text{FD} = 8.20$\,MB. In contrast, the  decentralized consensus-sharing (CS) equalizers in~\cite{li2017decentralized} require multiple iterations of message passing. Furthermore, in each each iteration the message passing data of both data gathering and consensus broadcasting doubles compared to that of the FD architecture, i.e., we have $m_\text{CS} = 2m_\text{FD} = 16.40$\,MB per iteration.}

\subsubsection{\revision{Practical Considerations}}
\revision{While our multi-GPU implementations exceed 1\,Gb/s throughput, the estimated total power dissipation of $C$ Tesla V100 GPUs with $300$\,W thermal design power (TDP) each can be as high as $C\times 300$\,W. To reduce the power consumption in practice, one can resort to FPGA or ASIC implementations. In fact, some of the latest massive MU-MIMO testbeds, such as the LuMaMi~\cite{lund,lund2017} testbed, are built with FPGA-driven software-defined radios and PXIe switches, and demonstrate 0.8\,Gb/s throughput for a $B=100$ antenna $U=12$ user system. However, the LuMaMi testbed processes the entire workload for different sub-bands (a small portion of the entire frequency band) on different FPGA co-processors, while each sub-band still connects to all BS antennas. Thus the PXIe switches can be saturated when further scaling up the antenna number. In contrast, the proposed feedforward   architectures divide the processing workload \emph{across antennas}, which yields improved scalability and modularity. In practice, one can combine antenna domain and frequency domain decentralization to combine the best of both approaches.}


   \section{Conclusions}\label{sec:conclusions}
   We have presented two feedforward architectures for decentralized equalization in massive MU-MIMO systems that mitigate the interconnect and I/O bandwidth bottlenecks and enable parallel processing on multiple computing fabrics.
   For the two proposed architectures, we have presented linear and nonlinear equalization algorithms, and we have analyzed their post-equalization SINR performance in the large-antenna limit. We have also performed numerical simulations that confirm our analysis.
   Our results indicate that nonlinear equalizers are able to achieve near-optimal SINR performance while enabling decentralized computations and low communication overhead among the antenna clusters. Linear equalizers perform equally well for scenarios in which the number of BS antennas is significantly larger than the number of UEs or for systems that use strong coding or low data rates. 
   Our reference implementations on a multi-GPU system have shown that our feedforward architectures achieve throughputs in the Gb/s regime, even for massive MU-MIMO systems with hundreds of antenna elements. Specifically, our measurement results show that feedforward architectures are able to overcome the latency limits of existing decentralized baseband processing schemes, such as the consensus-sharing methods proposed in~\cite{li2017decentralized}.

   There are many avenues for future work. First, an implementation of our algorithms and architectures on multi-FPGA or multi-ASIC platforms would demonstrate the full capabilities of our solutions. 
   \revision{Second, an investigation of antenna partitioning schemes and beamforming-based methods for directional channels, such as those experienced in millimeter wave (mmWave) systems, is part of ongoing research.}
   Third, a theoretical analysis of the precoding architectures and algorithms proposed recently in \cite{LJCS2018} for the massive MU-MIMO downlink is left for future work.
   %

\section*{Acknowledgments}
The work of C.~Jeon and C.~Studer was supported in part by Xilinx, Inc.~and by the US National Science Foundation (NSF) under grants ECCS-1408006, CCF-1535897,  CCF-1652065, CNS-1717559, and ECCS-1824379. 
The work of K.~Li and J.~R.~Cavallaro was supported in part by Xilinx, Inc.~and  by the US NSF under grants ECCS-1408370, CNS-1717218, and CNS-1827940, for the ``PAWR Platform POWDER-RENEW: A Platform for Open Wireless Data-driven Experimental Research with Massive MIMO Capabilities.''
The authors thank T.~Goldstein for discussions on DBP, and we acknowledge the hardware support of the DGX-1 multi-GPU systems at the Nvidia Technology Center (the PSG Cluster).

\appendices

\section{Derivations and Proofs}

\subsection{Derivation of \fref{alg:LAMA_alg}}\label{app:LAMAPD_derivation}

\fref{alg:LAMA_alg} builds upon the original LAMA algorithm \cite{JGMS2015conf}:
\begin{align*}
\bmz^t &= \bms^t + \bH^\Herm \bmr^t\\
\bms^{t+1} &= \mathsf{F}(\bmz^t,\No(1+\tau^t))\\
\tau^{t+1} &= \frac{\beta}{\No}\langle\mathsf{G}(\bmz^t,\No(1+\tau^t))\rangle \\
\bmr^{t+1} &=  \bmy - \bH \bms^{t+1} + \frac{\tau^{t+1}}{1+\tau^t}\bmr^t.
\end{align*}
We start with $\bmymrc = \bH^\Herm \bmy$ and $\bG = \bH^\Herm \bH$ and define $\phi^t = \frac{\No\tau^t}{\beta}$ so that $\frac{\tau^{t+1}}{1+\tau^t} = \frac{\beta\phi^{t+1}}{\No + \beta\phi^t}$. Then,
%
\begin{align*}
\bH^\Herm\bmr^{t+1} &=  \bH^\Herm \bmy - \bH^\Herm\bH\bms^{t+1} + \frac{\tau^t}{1+\tau^t} \bH^\Herm\bmr^t\\
&=  
\bmymrc - \bG \bms^{t+1} + \frac{\beta\phi^{t+1}}{\No + \beta\phi^t} \bH^\Herm\bmr^t.
\end{align*}
\fref{alg:LAMA_alg} follows by noticing that $\bmv^{t+1} = \bH^\Herm\bmr^t = \bmz^t - \bms^t$.

\subsection{Proof of \fref{lem:optimal_fusion}}\label{app:optimal_fusion}
As in \fref{eq:decoupling}, we write the estimate $z_{c,u}$ for UE $u$ at cluster $c$ as 
$z_{c,u} = s_{0u} + e_{c,u}$,
where $e_{c,u}$ represents residual interference and noise with known error variance $\Ex{}{|e_{c,u}|^2}=\sigma^2_{c,u}$. 
At UE~$u$, optimal fusion is $z_{u} = \sum_{c=1}^C \nu_{c,u} z_{c,u}$ so that \mbox{$\sum_{c=1}^C\nu_{c,u}=1$}. Hence, the fused estimate is 
$z_{u} = s_{0u} + \sum_{c=1}^C \nu_{c,u} e_{c,u}$,
with the following post-fusion SINR:
\begin{align} \label{eq:postfusionSINR}
 \SINR = \frac{\Ex{}{|s_{0u}|^2}}{\Ex{}{\big|\sum_{c=1}^C\nu_{c,u}e_{c,u}\big|^2}} = \frac{\Es}{\sum_{c=1}^C \nu_{c,u}^2 \sigma^2_{c,u}}.
\end{align}
Here, we used the assumption that the residual interference and noise terms $e_{c,u}$ are zero mean and uncorrelated across clusters $c=1,\ldots,C$. We are now interested in maximizing the post-fusion SINR in \fref{eq:postfusionSINR} subject to  $\sum_{c=1}^C\nu_{c,u}=1$. Using the method of Lagrange multipliers, it is easy to see that 
\begin{align}
\nu_{c,u} = \frac{1}{\sigma_{c,u}^2}
\left(\sum_{c'=1}^C \frac{1}{\sigma_{c',u}^2} \right)^{\!\!-1}, \quad c=1,\ldots,C.
\label{eq:opt_fusion}
\end{align}

\subsection{Proof of \fref{lem:SE_FP_Decentralized}}\label{app:SE_FP_Decentralized}
For Rayleigh-fading channels, each entry in the partial channel matrix $\bH_c$ is distributed as $\setC\setN(0,1/\MR)$.
To ensure that the expected column-norm of $\bH_c$ is one, we normalize the per-cluster input-output relation in \fref{eq:perclusterInOutrelation} by $1/\sqrt{w_c}$. This normalization amplifies the noise variance by $1/w_c$ in each cluster. In addition, since overall system dimension is $B w_c \times U$, the resulting system ratio is given by $\beta=U/(B w_c) = \beta/w_c$. 
By realizing that $\No/w_c$ is the per-cluster noise variance, the fixed-point equation follows immediately from Theorems \ref{thm:THeq} and~\ref{thm:SE} for linear and LAMA-based equalization, respectively.

\subsection{Proof of \fref{thm:postfusionvariance}} \label{app:postfusionvariance}

To simplify notation, we omit the UE index $u$. 
The proof follows from \fref{eq:opt_fusion} in \fref{lem:optimal_fusion}. 
The first expression in \fref{eq:sigma_FD} is trivial whereas the second expression is obtained as follows:
\begin{align*}
& 
\beta\sum_{c=1}^{\C} \nu_c \Psi(\bar\sigma_c^2) 
= 
\left(\sum_{c=1}^{\C}\frac{1}{\bar\sigma_c^2}\right)^{\!\!\!-1}\!
\sum_{c=1}^{\C} \frac{\beta\Psi(\bar\sigma_c^2)}{\bar\sigma_c^2}
\\ & \quad = 
\left(\sum_{c=1}^{\C}\frac{1}{\bar\sigma_c^2}\right)^{\!\!\!-1}\!
\sum_{c=1}^{\C}\left(w_c - \frac{\No}{\bar\sigma_c^2}
\right)
=
\left(\sum_{c=1}^{\C}\frac{1}{\bar\sigma_c^2}\right)^{\!\!\!-1}\!\!\!\! - \No.
\end{align*}

\subsection{Proof of \fref{lem:LAMA_Arch2_Arch1}}\label{app:LAMA_Arch2_Arch1}

We first show when equality holds. 
The case for $C=1$ is trivial because the \PD{} and \FD{} architectures are equivalent for $C=1$.
The case for $\beta \to 0$ is also straightforward because $\sigma_c^2=\sigma_\text{FD}^2=\sigma_\text{PD}^2 = \No$.
For MRC, we have $\sigma_\text{FD}^2 = \No+\beta\sum_{c=1}^{\C} \nu_c \Varop_S[S] = \No+\beta\Varop_S[S] = \sigma_\text{PD}^2$.

Let us now assume that $\beta>0$.
We show that $\sigma_c^2 > \sigma_\text{PD}^2$ by re-writing the fixed-point solutions as \cite{Maleki2010phd}:
$\sigma_c^2 = \sup\{ \sigma^2: \No+\beta\Psi(\sigma^2)\geq w_c\sigma^2\}$ 
and 
$\sigma_\text{PD}^2 = \sup\{ \sigma^2: \No+\beta\Psi(\sigma^2)\geq \sigma^2\}$.
Note that $\No>0$, so both $\sigma_c^2$ and $\sigma_\text{PD}^2$ are strictly positive.
It is easy to see that $\sigma_\text{PD}^2 \neq \sigma_c^2$ because $\sigma_\text{PD}^2  = \No+\beta\Psi(\sigma_\text{PD}^2 ) > w_c\sigma_\text{PD}^2$.
Since $\Psi(\sigma^2 ) \to \Varop_S[S]$ as $\sigma^2\to\infty$ and $\Psi(\sigma^2)$ is continuous \cite{GWSS2011}, there exists a $\sigma_c^2>\sigma_\text{PD}^2$ that satisfies $\No+\beta\Psi(\sigma_c^2)=w_c\sigma_c^2$ by the intermediate value theorem.
%

Finally, we use \cite[Prop. 9]{GWSS2011} to see that $\Psi(\sigma^2)$ is strictly increasing for $\sigma^2>0$ for LAMA. For ZF and MMSE, this also holds by inspection of $\dd \Psi(\sigma^2)/\dd\sigma^2 > 0$. 
Thus, the result $\sigma_\text{FD}^2 > \sigma_\text{PD}^2$ follows directly from \fref{lem:optimal_fusion} since
\begin{align*}
\sigma^2_\text{FD}  =\No \!+\! \beta \sum_{c=1}^{\C} \nu_c \Psi(\sigma_c^2) > \No \!+\! \beta \sum_{c=1}^{\C} \nu_c \Psi(\sigma_\text{PD}^2) 
= \sigma_\text{PD}^2.
\end{align*}

\subsection{Proof of \fref{lem:ZF_same}}\label{app:ZF_same}

The proof is straightforward and follows from \fref{thm:THeq} and \fref{lem:optimal_fusion}. 
Given that cluster $c$ has $B w_c > \beta$ antennas across all clusters $C$, the input-output relation of cluster $c$ in the large-system limit under ZF equalization results in a AWGN channel with decoupled noise variance:
%
${\sigma}_c^2  = \frac{\No}{ w_c - \beta }$.
%
The proof follows from \fref{lem:optimal_fusion} noting that $\sum_{c=1}^C w_c = 1$:
\begin{align*}
 \sigma_\text{FD}^2  = \left(\sum_{c=1}^C \frac{1}{{\sigma}_c^2}\!\right)^{-1} 
= \left(\sum_{c=1}^C \frac{w_c - \beta }{\No}\!\right)^{-1}
= \frac{\No}{ 1 - C\beta }.
\end{align*}

\subsection{Proof of \fref{lem:LMMSE_equalization}}\label{app:LMMSE_equalization}

The proof follows \fref{thm:THeq} with the fixed-point equation
\begin{align*}
 w_c \sigma_c^2 = \No + \beta \frac{\Es}{\Es+\sigma_c^2} \sigma^2_c,
\end{align*}
which results in the following \SINR expression for cluster $C$ for L-MMSE with the FD architecture:
\begin{align*}
 \SINR_{\textnormal{FD},c}^\textnormal{L-MMSE} =&\,\,  
\frac{1}{2}
\Bigg(
\sqrt{
\left(1 - \frac{\Es}{\No}(w_c-\beta)\right)^{\!\!2} + 4\frac{\Es}{\No}w_c}
\\
&- 
\Big(1 - \frac{\Es}{\No}(w_c-\beta)\Big)\!
\Bigg)\!.
\end{align*}
\fref{lem:LMMSE_equalization} follows from $\SINR_{\textnormal{FD}}^\textnormal{L-MMSE} = \sum_{c=1}^C \SINR_{\textnormal{FD},c}^\textnormal{L-MMSE}$.

\subsection{Proof of \fref{lem:LMMSE_upper}}\label{app:LMMSE_upper}
The proof of \fref{lem:LMMSE_upper} starts from \fref{eq:LMMSE_equalization}.
Let us denote $\overline{\SINR}_{\textnormal{FD}}^\textnormal{L-MMSE}$ as $\SINR_{\textnormal{FD}}^\textnormal{L-MMSE}$ when $w_1 = 0$ and $w_2=\cdots=w_C = 0$. We also define $\overline{\beta} = 1-\beta$.
Then,
\begin{align*}
&\!\!\!\!\!\!\!\!\!\!\!\max_{\bmw}\SINR_{\textnormal{FD}}^\textnormal{L-MMSE} 
\geq
\overline{\SINR}_{\textnormal{FD}}^\textnormal{L-MMSE}
\\=& 
\frac{1}{2}
\Bigg(
\sqrt{
\Big(1 - \frac{\Es}{\No}\overline{\beta}\Big)^{2} + 4\frac{\Es}{\No}}
- 
\Big(1 - \frac{\Es}{\No}\overline{\beta}\Big)\Bigg)
\\&+  
\frac{C-1}{2}
\Bigg(
\sqrt{
\Big(1 - \frac{\Es}{\No}\beta\Big)^{2}}
- 
\Big(1 - \frac{\Es}{\No}\beta\Big)
\Bigg)\!
\\
=& 
\frac{1}{2}
\Bigg(
\sqrt{
\Big(1 - \frac{\Es}{\No}\overline{\beta}\Big)^{2} + 4\frac{\Es}{\No}}
- 
\Big(1 - \frac{\Es}{\No}\overline{\beta}\Big)
\Bigg)
\\
\stackrel{(a)}{=}& 
\SINR_{\textnormal{PD}}^\textnormal{L-MMSE},
\end{align*}
where $(a)$ follows from (L-MMSE) in \fref{cor:sinr_PD}.
Since we know from \fref{lem:LAMA_Arch2_Arch1} that $\SINR_{\textnormal{PD}}^\textnormal{L-MMSE}\geq \SINR_{\textnormal{FD}}^\textnormal{L-MMSE}$, we have that $\max_{\bmw}\SINR_{\textnormal{FD}}^\textnormal{L-MMSE} = \overline{\SINR}_{\textnormal{FD}}^\textnormal{L-MMSE} = \SINR_{\textnormal{PD}}^\textnormal{L-MMSE}$.

\subsection{Proof of \fref{lem:LMMSE_uniform}}\label{app:LMMSE_uniform}
The proof of \fref{lem:LMMSE_uniform} starts from \fref{eq:LMMSE_equalization} with the definition $f(w,\alpha) = \sqrt{(1 - \alpha(w-\beta))^{2} + 4\alpha w}$. 
Note that $f(w,\alpha)$ is convex in $w$ as $f''(w,\alpha) \geq 0$ for $\alpha\geq 0$.
The final step follows from Jensen's inequality, which implies
\begin{align*}
 \frac{1}{C}\sum_{c=1}^C f(w_c,\alpha)  \geq f\left(\frac{1}{C}\sum_{c=1}^C w_c, \alpha\right),
\end{align*}
where equality holds if $w_1=w_2=\cdots=w_C$.

\balance

\bibliographystyle{IEEEtran}
\bibliography{bib/VIPabbrv,bib/confs-jrnls,bib/publishers,bib/VIP_180730}

\balance

\end{document}